# Extending the colloidal transition metal dichalcogenide library to $ReS_2$ nanosheets for application in gas sensing and electrocatalysis

*Beatriz Martín-García,[a,b,*] Davide Spirito,[c] Sebastiano Bellani,[a] Mirko Prato,[d] Valentino Romano,[a,e] Anatolii Polovitsyn,[b,f] Rosaria Brescia,[g] Reinier Oropesa-Nuñez,[h] Leyla Najafi,[a] Alberto Ansaldo,[a] Giovanna D'Angelo,[e] Vittorio Pellegrini,[a,g] Roman Krahne,[c] Iwan Moreels[b,f,*] and Francesco Bonaccorso[a,h,*]*

[a] Graphene Labs, Istituto Italiano di Tecnologia, via Morego 30, 16163 Genova, Italy.
[b] Nanochemistry Department, Istituto Italiano di Tecnologia, via Morego 30, 16163 Genova, Italy.
[c] Optoelectronics Group, Istituto Italiano di Tecnologia, via Morego 30, 16163 Genova, Italy.
[d] Materials Characterization Facility, Istituto Italiano di Tecnologia, via Morego 30, 16163 Genova, Italy.
[e] Dipartimento di Scienze Matematiche ed Informatiche, Scienze Fisiche e Scienze della Terra, Università di Messina, Viale F. Stagno d'Alcontres 31, S. Agata, 98166 Messina, Italy.
[f] Department of Chemistry, Ghent University. Krijgslaan 281-S3, 9000 Gent, Belgium.
[g] Electron Microscopy Facility, Istituto Italiano di Tecnologia, via Morego 30, 16163 Genova, Italy.
[h] BeDimensional Spa., Via Albisola 121, 16163 Genova, Italy.

Dr B. Martín-García, Dr S. Bellani, Dr L. Najafi, Dr A. Ansaldo, Dr V. Pellegrini, and Dr F. Bonaccorso
Graphene Labs, Istituto Italiano di Tecnologia, via Morego 30, 16163 Genova, Italy
E-mail: beatriz.martin-garcia@iit.it; francesco.bonaccorso@iit.it
Dr D. Spirito and Dr R. Krahne
Optoelectronics Group, Istituto Italiano di Tecnologia, via Morego 30, 16163 Genova, Italy
Dr M. Prato
Materials Characterization Facility, Istituto Italiano di Tecnologia, via Morego 30, 16163 Genova, Italy
V. Romano and Prof. G. D'Angelo
Dipartimento di Scienze Matematiche ed Informatiche, Scienze Fisiche e Scienze della Terra, Università di Messina, Viale F. Stagno d'Alcontres 31, S. Agata, 98166 Messina, Italy
Dr A. Polovitsyn and Prof. I. Moreels
Department of Chemistry, Ghent University. Krijgslaan 281-S3, 9000 Gent, Belgium
E-mail: iwan.moreels@ugent.be
Dr R. Oropesa-Nuñez, Dr V. Pellegrini, and Dr F. Bonaccorso
BeDimensional Spa., Via Albisola 121, 16163 Genova, Italy.
Dr R. Brescia
Electron Microscopy Facility, Istituto Italiano di Tecnologia, via Morego 30, 16163 Genova, Italy.

## Abstract

Among the large family of transition metal dichalcogenides (TMDCs), recently $ReS_2$ has stood out due to its nearly layer-independent optoelectronic and physicochemical properties.

These are related to its 1T distorted octahedral structure, which leads to strong in-plane anisotropy and the presence of active sites at its surface, which makes $ReS_2$ interesting for applications such as gas sensors and catalysts for $H_2$ production. However, the current fabrication methods for $ReS_2$ use chemical or physical vapor deposition (CVD or PVD) processes that are costly and involve complex and lengthy fabrication procedures, therefore limiting their large-scale production and exploitation. To address this issue, we developed a colloidal synthesis approach, which allows the production of $ReS_2$ to be attained at temperatures below 360 °C and with reaction times < 2 h, resulting in a more cost-efficient strategy than the CVD and PVD methods. By combining the solution-based synthesis with surface functionalization strategies, we demonstrate the feasibility of colloidal $ReS_2$ nanosheet films for gas sensing of different toxic gases, moisture and other volatile compounds with highly competitive performance in comparison with devices built with CVD-grown $ReS_2$ and $MoS_2$. In addition, the integration of the $ReS_2$ nanosheet films in assemblies, in which they are deposited on top of networks of carbon nanotubes, allowed us to fabricate electrodes for electrocatalysis for $H_2$ production in both acid and alkaline conditions. Results from proof-of-principle devices show an electrocatalytic overpotential that is competitive with devices based on $ReS_2$ produced by CVD, and even with $MoS_2$, $WS_2$ and $MoSe_2$ electrocatalysts.

## 1. Introduction

Transition metal dichalcogenides (TMDCs) are highly interesting and versatile materials due to their physicochemical properties that can be modified by exfoliation into single- or few-layered structures.[1–4] Due to quantum confinement and/or surface effects, such single and few-layered structures can behave differently from their bulk phase and manifest photoluminescence,[5] catalytically active sites[6] or strong light-matter coupling.[7] Moreover, depending on their composition and structure, they can be semiconductors (*e.g.* $MoS_2$, $WS_2$), metals (*e.g.* $NbS_2$, $VSe_2$) or even superconductors (*e.g.* $NbSe_2$, $TaS_2$).[1,8] Among the TMDCs, rhenium disulfide ($ReS_2$) has recently emerged as an interesting material.[9–11] In contrast to the well-known group-$VI_B$ TMDCs ($MoS_2$ or $WS_2$), which present a hexagonal crystalline structure, $ReS_2$ has an unusual distorted octahedral (1T) triclinic structure, leading to a strong in-plane anisotropy as the black phosphorous (BP).[9,10] However, unlike BP, $ReS_2$ is stable in air.[9,10] The $ReS_2$ structure consists of a $P\bar{1}$ symmetry with Re atoms forming $Re_4$ clusters, interconnected in zig-zag chains, and distorted $S_6$ octahedra that are formed upon cooperative atomic displacement (**Scheme 1**).[9],[11],[12,13] Moreover, $ReS_2$ exhibits both metal-chalcogen and metal–metal bonds due to the extra valence electron of Re atoms, since Re belongs to the group-VII of elements.[9,11–13] This structural anisotropy translates to electrical, optical, vibrational,[15] thermal[16] and physicochemical properties that do not depend significantly on the number of layers. For example, $ReS_2$ is a direct-gap semiconductor (1.4-1.5 eV) from bulk to monolayer (0.7 nm) thicknesses,[15] unlike $MoS_2$ or $WS_2$ that instead present an indirect to direct band gap ($E_g$) transition when the thickness is reduced from bulk to a monolayer.[9,10] The predictability of the physical properties of $ReS_2$ has made it attractive for a large variety of applications, in which it can be integrated as monolayer or few layer crystals in polarization-sensitive photodetectors,[17–19] field effect or heterojunction transistor structures,[20,21] and gas sensors,[25][26] or as thick nanocrystals or nanocrystal sheets with some tens up to hundred layers in batteries,[24–26] solar cells,[27] and electrocatalysts.[28,29] Despite its favorable properties, the feasibility of $ReS_2$ in technological and economic terms will strongly depend on the development of a scalable synthesis that can be readily integrated with the current device technology, being compatible with economic constrictions. In this regard, there are several challenges. First, Re is not an earth-abundant element, and therefore, its price (around 2800 €/kg[30]), determined by its availability and the market demand, is high in comparison to Mo and W (less than 30 €/kg[31]). Moreover, current methods for the production of $ReS_2$ are mainly based on chemical vapor deposition[19,32–34] (CVD), epitaxial growth[35] and the Bridgman method,[20] and rely on high process temperatures (from 450° to 1100°C) for the precursor decomposition (*e.g.*, the melting points of Re powder – 3180°C; Re–Te eutectoid 430°C; $ReO_3$ 400°C).[9,10] Moreover, they also imply the use of halogen vapor transport in the case of CVD[32]; or HF treatments for cleaning in the Bridgman method[20] and as etchant to delaminate the $ReS_2$ from the mica substrate in the epitaxial growth[35]. In addition, the aforementioned methods require long processing times from several hours for CVD[9,10] and epitaxial growth[9,10] up to several weeks for the Bridgman method.[9,10,20] Solution-based techniques such as chemical-intercalation[36] and liquid phase exfoliation (LPE)[37,38] have been developed as large-scalable routes for the $ReS_2$ flakes production, but in most cases CVD-grown $ReS_2$ is used as starting material, and therefore these methodologies inevitably still rely on expensive and demanding processing. To face the challenges related to the fabrication of $ReS_2$, colloidal synthesis can be an alternative method for its large scale production,[39–41] achieving a compromise between crystal quality and physical properties that are needed for device applications. In fact, colloidal synthesis has already been successful in the development of other TMDCs, such as $MoS_2$,[2,42–45] $MoSe_2$,[2,46,47] $MoTe_2$,[48] $WS_2$,[49] $WSe_2$,[46] $ZrS_2$,[50] $TiS_2$[50] and $HfS_2$[50].

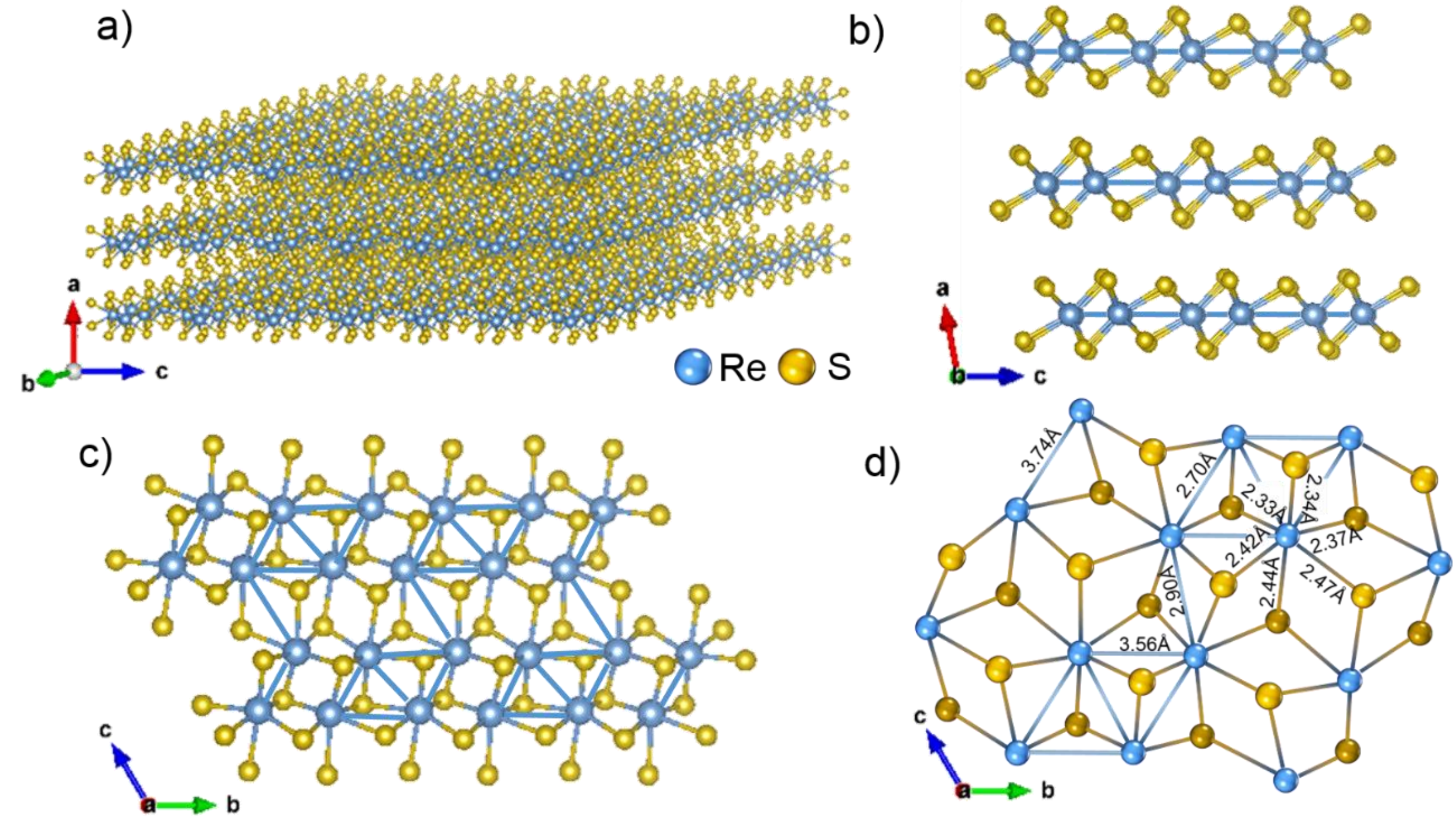

**Scheme 1.** Sketches of the distorted 1T crystal structure of the $ReS_2$ from different perspectives. Lattice orientations are indicated by the *a-b-c axes.* (a) A general view of a three-layer system, (b) side view in-plane of the layers; (c) top view of the three layers crystals where the $Re_4$ clusters can be identified, (d) enlarged top view indicating the distances between the atoms. To obtain the crystal structure we used the crystallographic data from refs. [12,13] and the VESTA *v.3.4.6* software[14].

Transition metal dichalcogenides have already been explored as gas sensing materials showing high sensitivity, room temperature operation, facile processing and high resistance to degradation, in order to address the main drawbacks of the materials that are currently used in environmental monitoring such as metal oxides[51], conducting polymers[51] and carbon nanotubes.[51] Recently, an interest in the use of $ReS_2$ for the fabrication of gas sensors[22,23] has emerged due to its strong interaction with non-metal adatoms (H, N, P, O, S, F, *etc.*). In transistors built from Scotch-tape exfoliated $ReS_2$,[22] a key role of the sulfur vacancies and interlayer interactions has been demonstrated to enhance the gas sensing capabilities, leading, for example, to better performance than $MoS_2$-[52] and graphene-based detectors[53] for $NH_3$ detection. In addition, $ReS_2$ is also sensitive to other gases such as $O_2$ or air, and can be applied in humidity sensors.[23]

The presence of active sites at the surface promoted by the stable and distorted 1T structure of the $ReS_2$ can be beneficial not only for the gas sensing application but also in electrocatalysis for the molecular hydrogen ($H_2$) production from electrochemical water splitting,[25,36,54] an important and growing field due to the high energy density of ~120-140 MJ $kg^{-1}$ and sustainability of $H_2$.[55–57] Actually, before $ReS_2$, other TMDCs have been reported as high-performance hydrogen evolution reaction (HER)-electrocatalysts.[6,58–60] In particular, the group-$VI_B$ TMDCs have reached an advanced stage of development,[6,58–60] exhibiting overpotentials inferior to 0.1 V at a cathodic current density of 10 mA $cm^{-2}$ ($\eta_{10}$) in both acid[61,62] and alkaline[63], [64,65] electrolytes. However, only the metallic edge states of their trigonal prismatic (2H) phase can absorb $H^+$ with a small Gibbs free energy ($\Delta G_H^0$ ~ 0.08 eV), acting as active site for HER,[66], [67,68] while the basal planes are electrocatalytically inactive.[67,68],[69] Recent advances have shown that the HER activity of group-$VI_B$ TMDCs can be significantly enhanced when the semiconducting 2H phase of $MoS_2$ is converted into metallic 1T (octahedral) phase.[70],[71] In this context, the stable and distorted 1T structure of the $ReS_2$ represents an interesting electrocatalyst model, which can be advantageous compared to group-$VI_B$ TMDCs due to its metal–metal bonds. These bonds create a superlattice structure of Re chains that distort the octahedral structure of 1T phase ($C_{3v}$ symmetry).[15,28] Consequently, the Gibbs free energy of the hydrogen adsorption ($\Delta G_H^0$) on $ReS_2$ basal planes can be as low as ~0.1 eV.[28] In fact, recent theoretical and experimental studies have shown that Re-Re bonds serve as electron reservoirs to originate an intrinsic charge distribution regulation, which tunes the Gibbs free energy of the hydrogen adsorption ($\Delta G_H^0$) on $ReS_2$ basal planes towards the optimal thermoneutral value (*i.e.*, 0 eV).[54] This value is comparable to that of metallic edge states of 2H-$MoS_2$ (~0.08 eV).[28,72] Thanks to the HER-activity of its basal planes, the $ReS_2$ should exhibit a higher number of active sites than group-$VI_B$ TMDCs.[28] In addition, in TMDCs such as $MoS_2$ or $WS_2$, the HER-activity strongly depends on the number of layers due to the dependence with the $E_g$, and the HER-activity weakens by passing from monolayer to few-layer materials as a consequence of an inefficient intra-flake electron transport *via* a hopping mechanism.[6][69] In contrast, the nearly layer-independent properties of $ReS_2$ make it a suitable catalyst, either monolayer or multilayer structures,[9,15] as it has been demonstrated in 3D structures composed by multilayer (≥17L) $ReS_2$ flakes[28,29].

In this work, we report a colloidal approach for the synthesis of $ReS_2$ nanosheets, starting from $ReCl_5$ and elemental sulfur as precursors. The obtained $ReS_2$ sheets are tested as active material in gas sensors and HER. For the colloidal synthesis of the $ReS_2$ nanosheets, we use elemental sulfur instead of $CS_2$[49][73][42] or dodecanethiol (DDT)[73][43] that are used as sulfur source for the colloidal synthesis of other TMDCs.[49][73][42][73][43] The advantage of using elemental sulfur relies on the fact that it is much less hazardous and/or toxic than the aforementioned precursors. We compare the properties of the colloidal $ReS_2$ with their bulk and LPE counterparts, and find that the colloidal $ReS_2$ nanosheets have a similar composition and $E_g$, but lower crystallinity. We fabricated gas sensors for a variety of agents by electrically contacting homogenous films of colloidal $ReS_2$ sheets. With suitable ligands for the surface functionalization of the colloidal $ReS_2$ nanosheets, we are able to enhance the gas sensor sensitivity to toxic gases ($NH_3$) and humidity, compared to their non-functionalized counterparts. Sensitivity, recovery and time-response of the functionalized gas sensors based on colloidal $ReS_2$ can compete with gas sensors fabricated from CVD-produced $ReS_2$ reported in literature.[23] Concerning the HER activity in acidic media, a mixture of colloidally synthesized $ReS_2$ in combination with carbon nanotubes, resulted in electrodes that can compete with the state-of-art CVD (KI)-aided $ReS_2$.[54]

## 2. Results and discussion

The colloidal $ReS_2$ sheets were synthesized from $ReCl_5$ and elemental sulfur by a syringe pump method,[49][47][42][48][44] in which the Re-precursor is added dropwise in a S-oleylamine (S-OlAm) solution at 350°C. This procedure follows the use of chloride metal precursors[49][73][42][48] and the progressive injection method[49][47][42][48][44] for the colloidal synthesis of TMDCs, but it introduces elemental S as the sulfur source. The chalcogen source can play a role in the growth kinetics, in particular in the formation of $H_2S$. Cheon's group demonstrated that sudden $H_2S$ influx, triggered by $CS_2$, favors the formation of nanodisks, while continuous $H_2S$ formation with DDT leads to flakes.[73] With elemental sulfur as chalcogen source we obtained nanosheets, and therefore assume slow $H_2S$ kinetics in the oleylamine medium, with a release of ~75% over a 3h heating period.[74] For the nucleation and growth of the sheets, we add the $ReCl_5$-oleic acid (OA) precursor dropwise, with a syringe pump, to the hot solution. High-angle annular dark-field scanning transmission electron microscopy (HAADF-STEM) images in **Figure 1a-b** (see also **Figure S1a** in **Supporting Information,** SI) show the formation of colloidal $ReS_2$ nanosheets (hereafter, c-$ReS_2$) with variable dimensions, formed by individual domains with a lateral size of ~4±1 nm. Moreover, some c-$ReS_2$ nanosheets are found lying perpendicular to the support film (**Figure 1c-d**). From these images a thickness of ~0.4±0.1 nm is estimated for individual c-$ReS_2$ nanosheets. This value is lower than the 0.7 nm reported for mechanically exfoliated monolayers in $ReS_2$.[75] The discrepancy can be attributed to the heavy strain building up in these highly anisotropic colloidal 2D structures, leading to in-plane expansion of the lattice and contraction along the out-of-plane direction.[76] The obtained c-$ReS_2$ nanosheets are comparable in morphology to other colloidal TMDC materials such as $MoS_2$,[42][43][45] $WS_2$[50] or $WSe_2$[46] reported in literature.

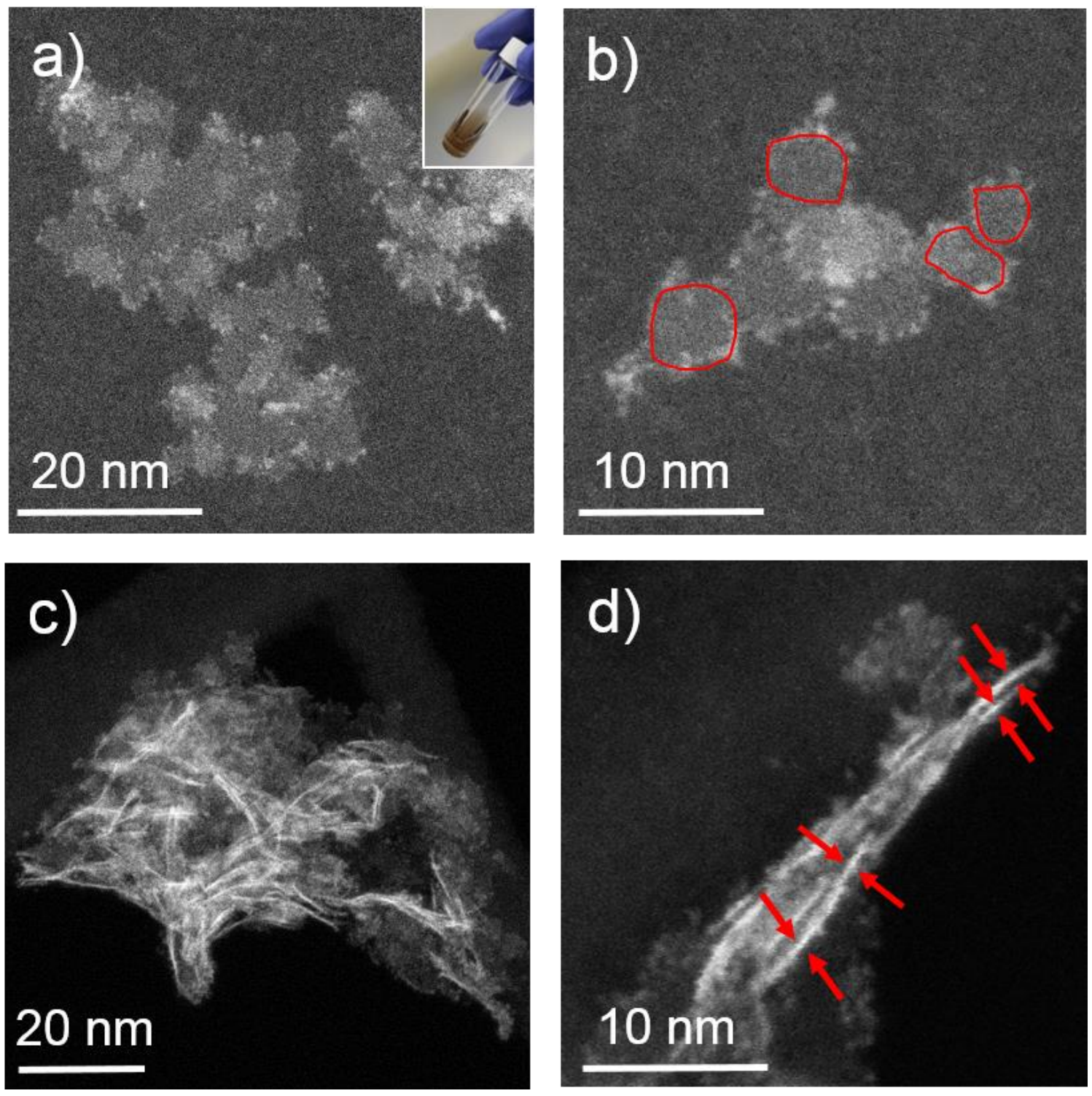

**Figure 1.** Representative HAADF-STEM images of the c-$ReS_2$ nanosheets, showing (a-b) nanosheets lying mainly parallel to the support film, with individual domains highlighted in (b). In (c,d) the nanosheets lie mainly perpendicular to the carbon support film, as clarified in (d) by arrows pointing at nanosheets viewed along their side. The inset in (a) shows a photograph of the c-$ReS_2$ dispersion in toluene.

The $ReS_2$ sheets prepared by LPE, as control material, do not show aggregation as demonstrated in the **SI**, **Figure S2a-b**. Regarding the nanosheet formation, previous studies in the CVD production of $ReS_2$ proved that there are two growth mechanisms: a fast one in (100) direction, and a slow one in (020) direction. To obtain nanosheets with a 1T distorted octahedral structure, the growth rate in both (100) and (020) directions should be comparable,[19] which may be achieved by oleic acid, oleylamine and $Cl^-$ (from the $ReCl_5$) acting as ligands that passivate different crystal facets. In order to gain insight into the crystal structure, we compare the selected-area electron diffraction (SAED) patterns from c-$ReS_2$ nanosheets and LPE-$ReS_2$ flakes showing comparable 2θ values for Bragg peak positions (see **SI**, **Figure S3** for the TEM-SAED patterns), which correspond to the 1T distorted octahedral structure. Furthermore, broader diffraction features characterize the c-$ReS_2$ nanosheets, due to the much smaller size of single-crystal domains (few nm or less in the colloidal sample *vs.* hundreds of nm for LPE flakes). The composition of the as-prepared c-$ReS_2$ was investigated by X-Ray photoelectron spectroscopy (XPS), as shown in **Figure 2a-b**, revealing an atomic stoichiometry Re:S of 1:1.3, which is estimated from the Re 4f and S 2p spectra. This ratio is slightly smaller than would be expected from the 1:2 $ReS_2$ stoichiometry. However, also the LPE fabricated samples and bulk $ReS_2$ yield lower Re:S stoichiometric values of around 1:1.6 and 1:1.7, respectively (see also **SI**, **Figure S1e** for complementary elemental analysis). It is interesting to note that in the LPE and bulk cases, the S 2p signal can be decomposed in two different doublets, corresponding to two different chemical environments of sulfur. Similar results have been reported in Ref. [77], and the two S components have been assigned to $ReS_2$ (doublet at lower binding energy, with S $2p_{3/2}$ component at approx. 161.4 eV) and to S atoms that are not connected to Re atoms (doublet at higher binding energy, with S $2p_{3/2}$ component at approx. 162.1 eV). However, we cannot discard that the two S components come from the existence of different Re-S bond lengths in the distorted $ReS_2$ structure, as illustrated in Scheme 1d.[12,13] For colloidal $ReS_2$, a good fitting of the S 2p profile was obtained by using two doublets as for the LPE and bulk-counterparts, with S $2p_{3/2}$ components at ∼161.6 eV and ∼162.2 eV. Moreover, XPS peaks in c-$ReS_2$ are broader than the ones for the LPE and bulk materials likely indicating a less crystalline structure in the colloidal case.[78] To support this statement, we also evaluated the extinction coefficient and Raman spectra that are plotted in **Figure 2c** and **d-e**. As expected for $ReS_2$,[9,10][79] the material shows a strong and broad extinction from 300 nm to almost 1000 nm. However, in contrast to its LPE equivalent, there is no clear excitonic peak at ~810 nm that corresponds to the $E_g$ of the $ReS_2$. [79] The $E_g$ was obtained from the $(\alpha h\nu)^n$ *vs.* $h\nu$ (Tauc plot) analysis (see **Figure S1d** and **Figure S2c** in **SI**) using the Tauc relation $Ah\nu = Y(h\nu\text{-}E_g)^n$, in which A is the absorbance, $h$ is Planck's constant, ν is the photon's frequency, and Y is a proportionality constant.[80] The value of the exponent denotes whether it is a direct transition ($n$ = 2) or an indirect one ($n$ = 0.5).[81] Since $ReS_2$ should be a direct bandgap semiconductor, we applied $n$ = 2, resulting in an $E_g$ of ∼1.41 eV (colloidal) and ∼1.43 eV (LPE). Both values are in agreement with the theoretical (1.41 eV)[82] and experimental values (1.4-1.5 eV) from bulk to monolayer (0.7 nm) thicknesses.[15] The analysis of the Raman spectra in **Figure 2d-e** has been performed following the detailed reports on $ReS_2$ carried out by Balicas' and Terrones' groups,[83,84] which identified 18 first-order modes in the 100-450 $cm^{-1}$ range (see **SI** for Raman peaks interpretation). Since in the randomly oriented assembly of the nanosheets we cannot control the crystal orientation, we performed a comparative Raman analysis with excitation at two different wavelengths, *i.e.*, 532 and 785 nm. Independently of the excitation wavelength and synthesis batch, the Raman peaks from the colloidal sample are broader than the corresponding ones in the LPE sample. Such broader linewidth in the Raman signal of the colloidal sample indicates a lower degree of crystallinity compared to the LPE one,[85] which can be related to the difference in the temperature that is needed for their preparation. Highly crystalline $ReS_2$ fabricated by CVD requires a temperature > 600°C,[35] while the colloidal synthesis is performed at 350°C. The low crystallinity of the colloidal $ReS_2$ nanosheets might also be the cause for the absence of the excitonic peak in the spectrum in Figure 2c.

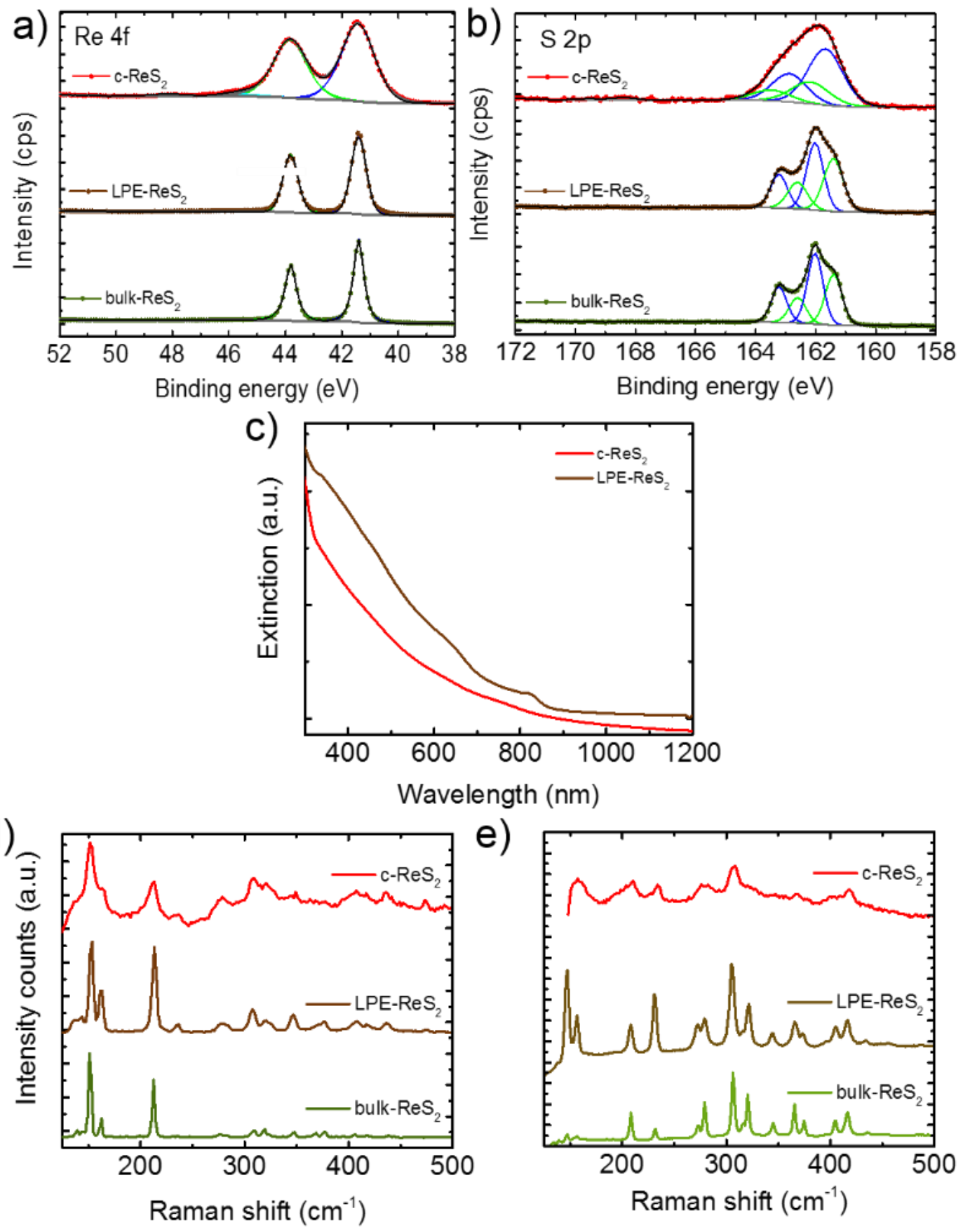

**Figure 2.** (a, b) Re 4f and S 2p XPS core-level spectra collected from the c-$ReS_2$, LPE-$ReS_2$ and bulk-$ReS_2$ samples, in which the fitting reveals the deconvoluted peaks. (c) Extinction spectra of c-$ReS_2$ and LPE-$ReS_2$ dispersed in toluene and isopropanol, respectively. An offset was applied to the curves for clarity. (d-e) Representative Raman spectra collected at (d) 532 and (e) 785 nm excitation wavelength on the c-$ReS_2$, LPE-$ReS_2$ and bulk $ReS_2$ samples on silicon wafers.

For both colloidal and LPE samples, the following Raman modes are enhanced at 785 nm excitation compared to the 532 nm excitation, $A_g^9$ (273 $cm^{-1}$), $A_g^{10}$ (279 $cm^{-1}$), $A_g^{12}$ (305 $cm^{-1}$), $A_g^{13\text{-}14}$ (322 $cm^{-1}$), $A_g^{16}$ (366 $cm^{-1}$), $A_g^{17}$ (373 $cm^{-1}$), $A_g^{18}$ (405 $cm^{-1}$) and $A_g^3$ (417 $cm^{-1}$), while the $A_g^{15}$ (350 $cm^{-1}$) and $A_g^2$ (440 $cm^{-1}$) modes disappear, indicating that the samples are composed by few-layered (2L-4L) $ReS_2$, in agreement with the data reported by Terrones' group.[84] In order to prove that our colloidal synthetic approach can be versatile also for other materials, we fabricated $ReSe_2$ using Se powder as chalcogen source (see **SI**, **Figure S4** for synthesis details and material characterization).

In order to test the suitability of the $ReS_2$ sheets as films for gas sensing, and to compare the performance of the colloidal and LPE samples, we prepared films on glass substrates by drop-casting (**Figure 3a**). For the c-$ReS_2$ nanosheets, we used the original dispersion for drop-casting, followed by annealing at 300°C in a glovebox (3 h) to remove the organic ligands from synthesis (oleic acid and oleylamine) with the aim to make the film electrically conductive. Films from LPE samples were fabricated by drop-casting without any further treatment. In this way, we obtained continuous and compact films with thickness of ~400 nm, as shown in the SEM images in **Figure 3b-c** (see **Experimental Section** for fabrication details). The devices were completed by the evaporation of Au electrodes as contacts. All devices manifest ohmic behavior (see **SI**, **Figure S5a** for conductance measurements under inert -$N_2$- atmosphere), however the resistance of the devices made from the colloidal $ReS_2$ is much lower (460 MΩ) than the one of devices fabricated with LPE-$ReS_2$ (5.9 GΩ). For gas sensing *via* the detection of conductance changes of the film, we positioned the samples inside a chamber with controlled atmosphere connected to a probe station (see **Figure 3a** and the **Experimental Section** for details on the gas filling of the chamber and measurement protocols). From the gas flow rates we estimate the filling and purging time of the chamber to be around 1s, which is shorter than the response times that we measured on our devices. We tested the response to humidity, to $NH_3$ that is a toxic agent, and to ethanol (EtOH) and acetone as representative volatile organic compounds. **Figure 3d-f** shows the representative response ($\mathbb{R}$) for different gases of the devices obtained *via* the relative changes of the film conductance (G)[51] in the presence of the target gas compared to inert ($N_2$) atmosphere, $\mathbb{R} = \left(G_{gas} - G_{inert}\right)/G_{inert}$. The numerical values are reported in the **SI** in **Table S1**, in which also the response in terms of resistance is given for comparison with literature.

For $NH_3$ and $H_2O$ the response of the gas sensors is positive, *i.e.* the conductance under the target gas is higher compared to that under the inert $N_2$ atmosphere. This can be explained by physisorption of the gas molecules that act as electron donors.[22,23,51,86,87] The higher response for gas sensors based on $ReS_2$ under $NH_3$ compared to air and $O_2$ was also reported in theoretical studies.[22] For $NH_3$ detection, the relative response of the devices built from LPE-$ReS_2$ flakes (hereafter, 'LPE devices') (Figure 3f) is much higher compared to the ones fabricated from annealed c-$ReS_2$ nanosheets (in the following noted as 'colloidal devices'). However, this result is mainly related to the very low electrical conductivity of the LPE device under inert atmosphere. In fact, the recovery time of the LPE devices to the initial currents (*i.e.*, in the nA range) is of the order of hours, while the recovery time of the colloidal devices is of the order of seconds (see **SI**, **Table S2** for rise and fall times of conductance response of the different devices; and **Tables S3** and **S4** for estimated minimum amounts detectable with the technique). In practical terms this means that for detection with response times of few seconds the baseline of the LPE device is at a value of ~1000, and therefore the relative response to $NH_3$ at such time scales is ~10, while that of the colloidal device is ~2.5 - 3. Concerning humidity detection, the response of the devices made from colloidal $ReS_2$ ($\mathbb{R}$ = 1.2; rise time ($\tau_R$ = 8.7 s); fall time ($\tau_F$ = 0.3s)) outperforms clearly the ones based on LPE flakes ($\mathbb{R}$ = 8.8; $\tau_R$ = 5.2 s; $\tau_F$ = 31s). In fact, the recovery time to return to 10% of the response after the gas flow stopped is one order of magnitude shorter for c-$ReS_2$ compared to the LPE flakes one. For EtOH, acetone and compressed air the conductance of the devices made from c-$ReS_2$ decreases compared to one achieved in inert $N_2$ atmosphere (**Figure 3e**). This result can be rationalized by oxidation of the film due to the incorporation of electron acceptor molecules.[51,86,87] Furthermore, the relative conductivity change of the devices made from c-$ReS_2$ exposed to EtOH, acetone and compressed air is much weaker ($\mathbb{R}_{EtOH}$ = -0.15; $\mathbb{R}_{acetone}$ = -0.12; and $\mathbb{R}_{air}$ = -0.06) and slower in recovery (EtOH: $\tau_R$ = 5.2 s; $\tau_F$ = 25s; Acetone: $\tau_R$ = 1 s; $\tau_F$ = 56s; and air $\tau_R$ = 2.3 s; $\tau_F$ = 12s) as compared to their exposure to $NH_3$ ($\mathbb{R}$ = 2.4; $\tau_R$ = 9.3 s; $\tau_F$ = 1.3 s) or $H_2O$ ($\mathbb{R}$ = 1.2; $\tau_R$ = 8.7 s; $\tau_F$ = 0.3 s).

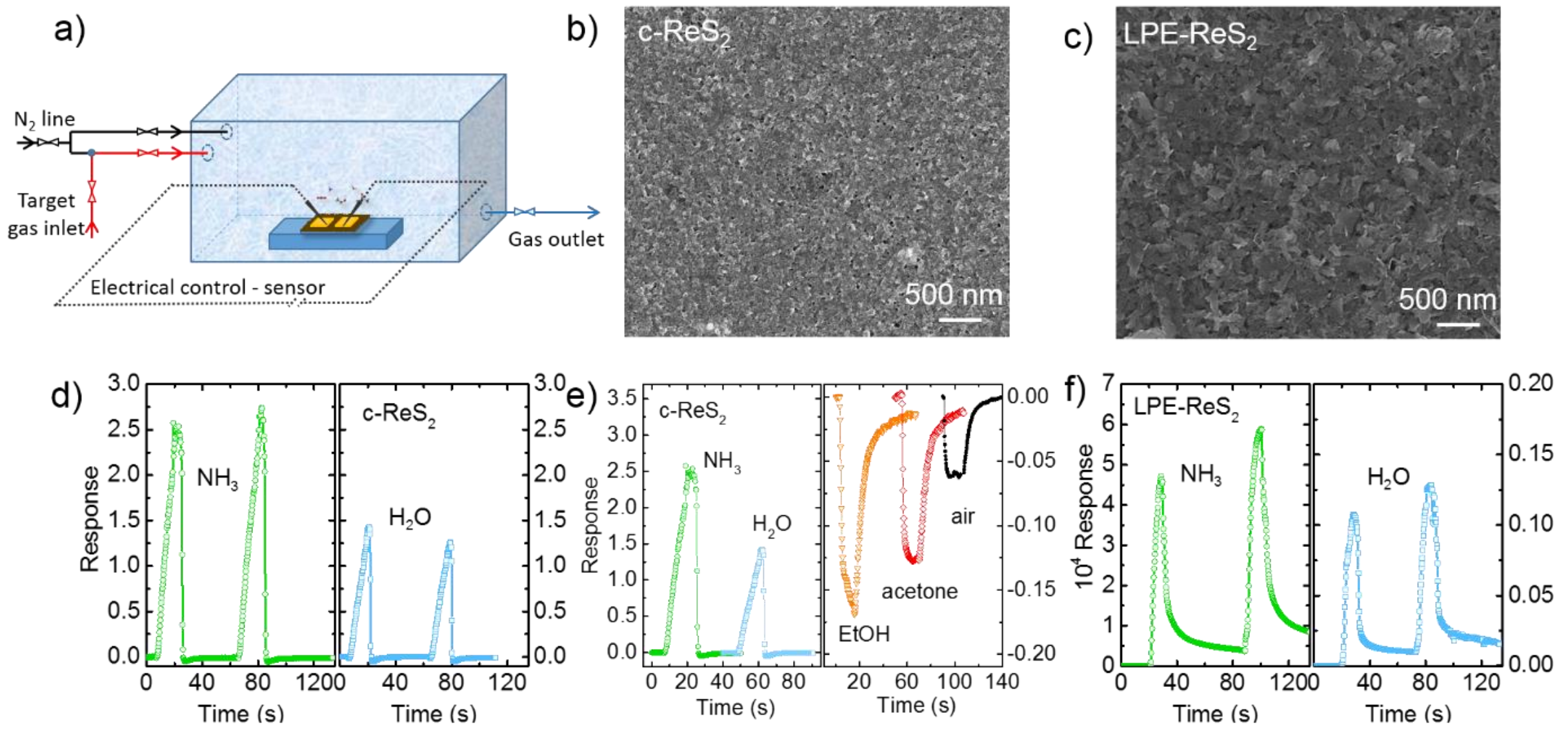

**Figure 3.** (a) Sketch showing the chamber used for the gas sensing experiments. (b,c) Representative SEM images of the colloidal and LPE-$ReS_2$-based gas sensors, respectively. The films (~400 nm thickness) were prepared by drop-casting the corresponding dispersions on glass substrates. (d) Representative gas induced-time response of the device built with ann-$ReS_2$ in two consecutive cycles exposed to $NH_3$ (left) and $H_2O$ (right). (e) Representative gas induced-time response of the device built with ann-$ReS_2$ exposed to different gases: $NH_3$, $H_2O$ (left), EtOH, acetone and dry air (right). (f) Representative gas induced-time response of the device built with LPE-$ReS_2$ exposed in two consecutive cycles to $NH_3$ (left) and $H_2O$ (right). In all the cases, the gas-induced response was determined from the film conductance variation as detailed in the text.

The circumstance that the colloidal $ReS_2$ nanosheets are passivated by organic ligands opens the possibility to manipulate the film conductivity by ligand exchange with other molecules. We therefore use this approach to improve the performance of the c-$ReS_2$ based sensors, in terms of recovery time and response. In particular, we performed a ligand exchange process that replaces the long-chain organic ligands used in the synthesis (oleic acid and oleylamine, both with a $C_{18}$ aliphatic chain) with shorter molecules that contain $C_x$ aliphatic chains with $x<4$ or an aromatic ring (see **Experimental Section** for more details). Shorter ligands lead typically to stronger coupling of the nanomaterial in a compact film, which increases the charge carrier mobility.[88,89] Moreover, chemical modification of the nanostructure surface by ligand exchange can modify the electronic properties of the material, as well as their reactivity with other functional groups such as gas molecules. [51,86] We tested different short chain molecules in this respect: 3-mercaptopropionic acid (MPA), 1,4-benzenedithiol (BDT) and 4-aminobenzoic acid (ABA), and performed the ligand exchange in solution. The films from the ligand exchanged solutions were prepared with the protocol described before, which ensured a compact film with a thickness of ~400 nm (**Figure 4a-b**, see **Figure S6** in the **SI** for additional characterization). A larger S contribution in the XPS signal (compared to the samples before ligand exchange) that results from the thiol groups of the exchanged molecules confirmed the presence of MPA and BDT molecules. Concerning the exchange with ABA, the success of the ligand exchange process can only be assessed from the increase of the conductivity, since the characteristic N 1s XPS peak ($NH_2$, at 400.5 eV) also originates from oleylamine. The ligand-exchanged films also show ohmic conductivity, and their resistance strongly depends on the individual ligands (see SI, **Figure S5b** for conductance measurements under inert -$N_2$- atmosphere). Here, devices made from MPA-exchanged c-$ReS_2$ nanosheets manifested the highest conductance (R = 120 MΩ), followed by ABA (R=510 MΩ) and BDT (R = 200 MΩ). Compared to the original device discussed in **Figure 3**, we achieved an increase in electrical conductivity by a factor of 4 with the MPA ligands. We tested the ligand-exchanged films for gas sensing with the same procedure as described before. Overall, we observe similar behavior before and after the ligand exchange process, with an electrical current increase under $NH_3$, $H_2O$ and $CO_2$, and, instead, decrease under EtOH, acetone and air **Figure 4c** depicts the response of the device built from MPA-exchanged c-$ReS_2$ nanosheets to different gases: $NH_3$, $H_2O$ and $CO_2$ that lead to current increase, and EtOH, acetone and air that result in a small current decrease. Therefore, we can assume that in all the cases, colloidal, LPE and ligand exchange fabricated devices, the sensing mechanism is based on a charge transfer between the physisorbed gas molecules and the $ReS_2$ structure.[51,87] **Figure 4d** shows the response of the gas sensors made with ligand-exchanged c-$ReS_2$ nanosheet films when exposed to $NH_3$. In this context, the device made from MPA-exchanged c-$ReS_2$ nanosheets shows the highest response ($\mathbb{R}$=31, with a response time of 3.3 s), followed by the ones built from BDT- and ABA-exchanged c-$ReS_2$ nanosheets. The functionalization of c-$ReS_2$ nanosheets with BDT yields a higher sensitivity ($\mathbb{R}$=24 response for $NH_3$ detection) than ABA ($\mathbb{R}$=16 response for $NH_3$ detection), although the conductance of ABA ($R_{ABA}$=510 MΩ vs $R_{BDT}$=200 MΩ) is higher under inert atmosphere. The different results in terms of sensitivity and responsivity obtained without and with the functionalization of c-$ReS_2$ nanosheets, and even more, by varying the molecules used for the functionalization, point to the importance of: (i) possible sites for the binding of gas molecules; (ii) the efficiency of charge transfer for the gas sensing sensitivity, and (iii) a good film conductivity, which is beneficial for a fast response and recovery. Here, we can expect a trade-off between a film with high surface area that is beneficial for interaction with the gas molecules, and a compact film that results in high conductivity and from which a fast response can be expected. In particular, the high sensitivity obtained with the BDT and MPA ligand exchange for $NH_3$ ($\mathbb{R}_{BDT}$=24 and $\mathbb{R}_{MPA}$=31, respectively) and $H_2O$ ($\mathbb{R}_{BDT}$=2.4 and $\mathbb{R}_{MPA}$=4.5, respectively) points to a beneficial role of the SH groups of MPA and BDT for physisorption of the gas molecules and the electron transfer towards the $ReS_2$ film. We note that the performance of the device made from MPA-exchanged c-$ReS_2$ nanosheets is superior to the one built with LPE-$ReS_2$ in **Figure 3** in terms of sensitivity and response time (*e.g.* for $NH_3$ detection, $\mathbb{R}$=9.9, with a response time of 5.2 s for LPE-$ReS_2$ and $\mathbb{R}$=31, with a response time of 3.3 s for MPA-exchanged c-$ReS_2$).

The gas sensor built from MPA-exchanged c-$ReS_2$ outperforms, to the best of our knowledge, those reported in literature fabricated with CVD-grown $ReS_2$.[23] The humidity sensor reported by Yang *et al.*[23] made by CVD-grown $ReS_2$ has a resistance variation of ~-60% at a relative humidity of 70 %, with a response time of the order of tens of seconds, while our gas sensor fabricated with MPA-exchanged c-$ReS_2$ reaches a variation of 80% with much faster response time (4.1 s) and comparable recovery time (1 s). Gas sensing ($NH_3$, $O_2$, air) has been also reported from phototransistors made from micro-mechanically exfoliated $ReS_2$ flakes,[22] with a response time in the order of ms. However, this response time is related to the illumination of the device and therefore does not directly compare to our devices that work in dark condition. Concerning the gas sensors for $NH_3$ and EtOH based on other 2D materials, we note that our MPA-functionalized sensor has a faster response time and better recovery of the initial conductance state than devices that use micro-mechanically exfoliated or CVD-grown $MoS_2$.[90] [91]

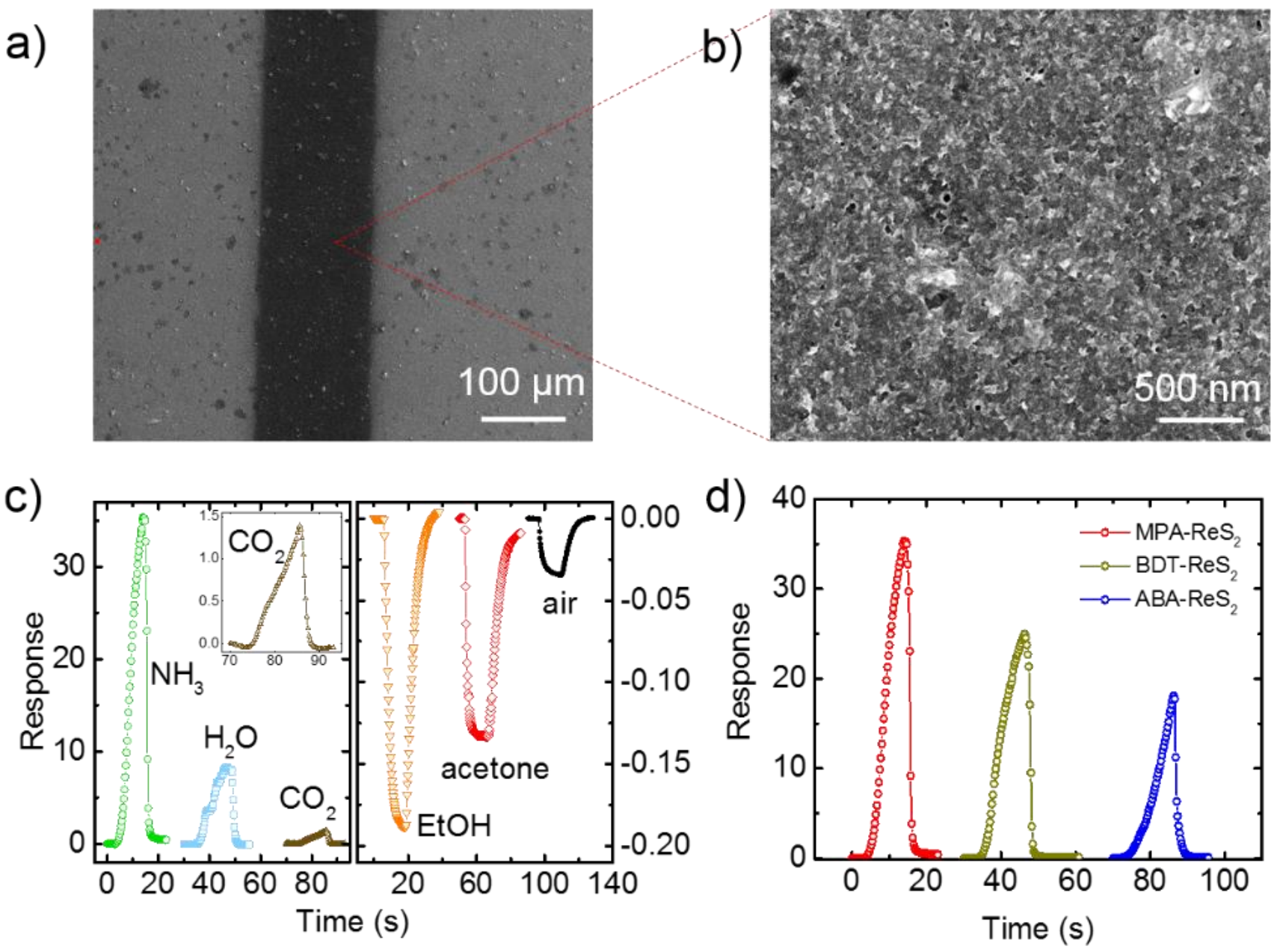

**Figure 4.** (a, b) Representative SEM images of the device built from MPA-exchanged c-$ReS_2$. The film (with ~ 400 nm film thickness) was prepared by drop-casting the corresponding dispersion. (c) Representative gas induced-response time of the device made from MPA-exchanged c-$ReS_2$ under the exposure to different gases: (left) $NH_3$, $H_2O$, and $CO_2$, (right) EtOH, acetone and dry air. The inset in the left panel shows the $CO_2$ data on a magnified scale. (d) Comparison of the device performance in terms of response to $NH_3$ exposure for the different ligand molecules used: MPA, ABA and BDT. In all the cases, the gas-induced response was determined from the film conductance variation as detailed in the text.

Finally, we tested c-$ReS_2$ films as an electrocatalytic film for HER (**Figure 5a**). To make the c-$ReS_2$ more electrochemically active, we isolated the nanosheets from the dispersion by centrifugation and solvent evaporation. The obtained powder was then treated by thermal annealing (Ar flow, 500 °C) to remove the organic ligands from synthesis, and thereby expose the catalytically active surface of the nanosheets. After the annealing, the $ReS_2$ powder was dispersed in N-methyl-2-pyrrolidone (NMP) by ultrasonication, and the obtained dispersion was used for the deposition on glassy carbon (GC) rigid electrodes and single-walled carbon nanotubes (SWCNTs)-based flexible papers (*i.e.*, buckypapers) (**Figure 5b**) (see **Experimental Section** for more details). The choice of the SWCNTs-based paper as support relies on our recent works,[61],[63] in which we demonstrated a long-range (≥ 1 µm) electrochemical coupling between HER-active TMDCs and SWCNTs for increasing the HER-activity of TMDCs. [61,63] Moreover, the porosity of such substrate promotes the adhesion of the TMDC films without the need of ion conducting catalyst binders[61],[63],[92][93](*e.g.*, Nafion in acid solution[94] and Tokuyama AS-4 in the alkaline one[95]), which can be detrimental to the electrocatalytic activity of the catalyst.[96],[97] The GC-based electrodes were obtained by drop-casting the c-$ReS_2$ dispersions at a mass loading of 0.13 mg cm$^{-2}$. The hybrid SWCNTs/c-$ReS_2$ electrodes were produced through a sequential vacuum filtration deposition of the material dispersions onto nylon membranes (material mass loading of 1.5 mg cm$^{-2}$ for both SWCNTs and c-$ReS_2$ (5 mg of each material)). **Figures 5c-d** display top-view and cross-sectional SEM images of a representative SWCNTs/c-$ReS_2$ electrode. The c-$ReS_2$ nanosheets form a film atop the SWCNTs network, and the electrode shows a bilayer architecture with ~20 µm-thick SWCNT–based collector and a thin c-$ReS_2$-based active film (thickness in the order of 1 µm). The high-magnification SEM image (**Figure 5c**) reveals the presence of $ReS_2$ aggregates with various dimensions, in the 0.5-10 µm range. The smallest $ReS_2$ aggregates (lateral size < 500 nm) penetrate the mesoporous SWCNT network (see **SI**, **Figure S9**). For such configuration, SWCNTs increase the electron accessibility to the HER-active sites of c-$ReS_2$ nanosheets, speeding up the HER-kinetics compared to flat GC.[98][61][92][99] **Figure 5e** shows a top-view photograph of the as-produced SWCNTs/c-$ReS_2$ (electrode area of ~3.5 $cm^2$), in which the presence of $ReS_2$ aggregates is visible by eye. Such aggregates are also evident in SEM images (**Figure 5f**) with a lateral size ranging from 50 to 200 μm.

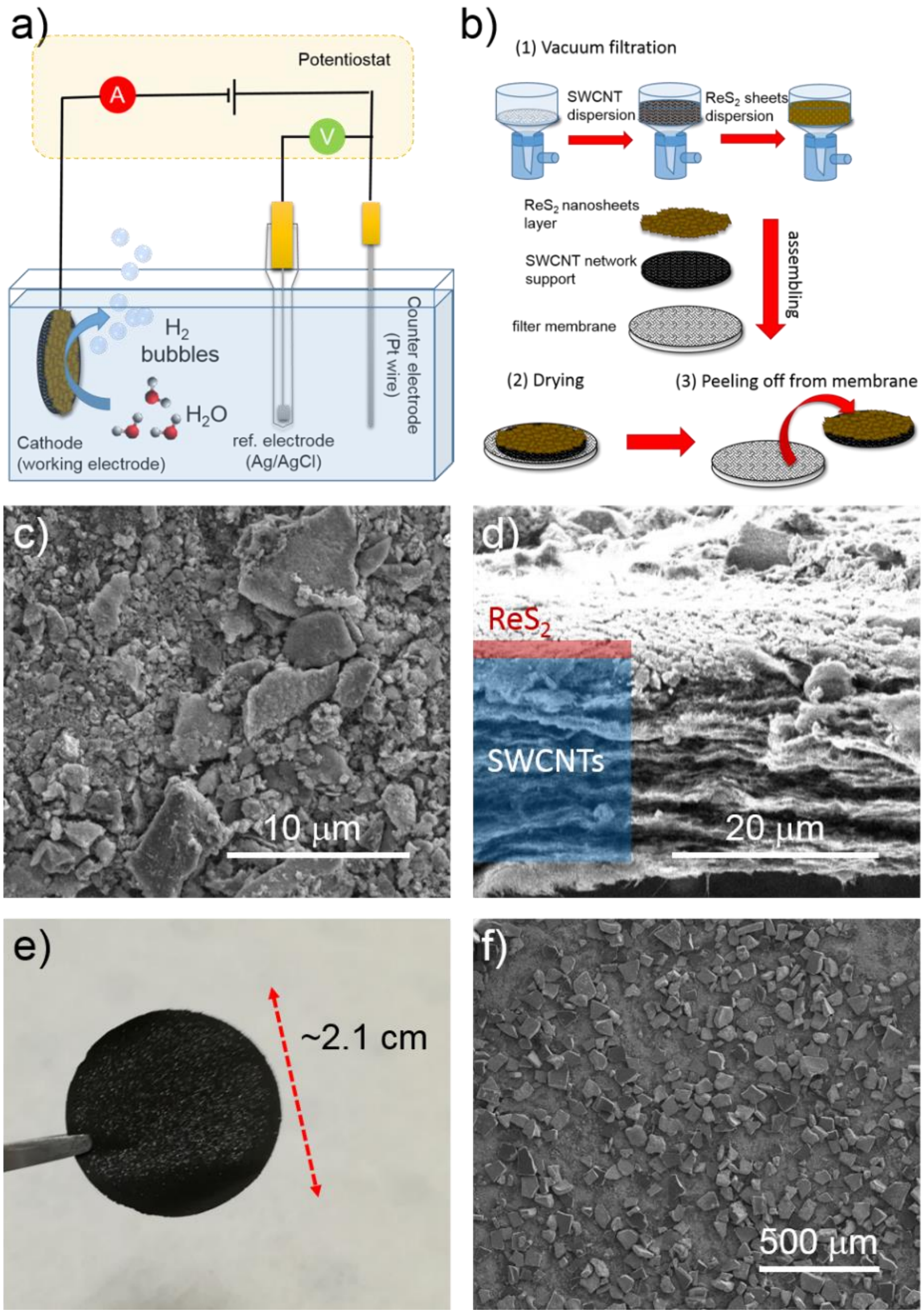

**Figure 5.** (a) Scheme of the electrochemical setup for measuring the electrocatalysts. (b) Sketch of the fabrication of the c-$ReS_2$-based electrodes on SWCNTs (c, d) Top-view and cross-sectional SEM images of SWCNTs/c-$ReS_2$. (e) Top-view photograph of SWCNTs/c-$ReS_2$ (electrode area: 3.5 $cm^2$). (f) Top-view SEM image of SWCNTs/c-$ReS_2$ at higher magnification compared to the one shown in panel c (scale bar: 500 μm).

The HER-activity of the SWCNTs/c-$ReS_2$ was tested in both in acidic (0.5 M $H_2SO_4$) and alkaline (1 M KOH) media. In principle, the HER in acidic solution proceeds with an initial discharge of the hydronium ion ($H_3O^+$) and the formation of atomic H adsorbed on the electrocatalyst surface ($H_{ads}$), in the so-called Volmer step ($H_3O^+ + e^- \rightleftarrows H_{ads} + H_2O$), followed by either an electrochemical Heyrovsky step ($H_{ads} + H_3O^+ + e^- \rightleftarrows H_2 + H_2O$) or a chemical Tafel recombination step ($2H_{ads} \rightleftarrows H_2$). In alkaline media, the $H_{ads}$ is formed by discharging $H_2O$ ($H_2O + e^- \rightleftarrows H_{ads} + OH^-$). Then, either a Heyrovsky step ($H_2O + H_{ads} + e^- \rightleftarrows H_2 + OH^-$) or a chemical Tafel recombination step ($2H_{ads} \rightleftarrows H_2$) occurs. Apart from the cathodic current density $\eta_{10}$, the Tafel slope is also an important Figure of Merit to assess the HER-activity.[100] However, a rigorous kinetic analysis of the HER establishing the Tafel slope was not carried out for our electrodes, since SWCNTs hold a high surface area that leads to a remarkable capacitive current density (in the order of 1 mA $cm^{-2}$) even at a low LSV sweep voltage rate (≤5 mV $s^{-1}$). This can be the cause of misleading interpretations of the estimated kinetic parameters.[63],[101]

**Figure 6a-b** show the current-resistance (iR)-corrected polarization curves of GC/c-$ReS_2$ and SWCNTs/c-$ReS_2$ electrodes in acidic (0.5 M $H_2SO_4$) and alkaline (1 M KOH) aqueous solutions, respectively. The series resistance that arises from the electrical resistance of the working electrode and the electrolyte resistance, *i.e.*, R, was extrapolated by single frequency electrochemical impedance spectroscopy (EIS) (see Experimental Section for further details). The curves measured for the substrates, *i.e.*, GC and SWCNTs, are also shown as references. The curve measured for commercial platinum on carbon (Pt/C) is reported as benchmark for HER. The c-$ReS_2$-based electrodes show an enhanced HER-activity compared to those of the substrates, *i.e.*, GC and SWCNTs. Noteworthy, the use of SWCNTs as the substrate significantly increases the cathodic current densities, *i.e.*, the HER-activity, compared to the one shown by CG/c-$ReS_2$. Thus, in acidic media, $\eta_{10}$ decreases from 0.465 V for the GC/c-$ReS_2$ to 0.196 V for SWCNTs/c-$ReS_2$. In alkaline solutions, GC/c-$ReS_2$ electrode is poorly HER-active ($\eta_{10}$ > 0.7 V). Interestingly, SWCNTs/c-$ReS_2$ electrode exhibits a $\eta_{10}$ of 0.299 V, indicating that the SWCNTs and c-$ReS_2$ nanosheets interact to synergistically enhance the HER-activity of the electrode. This effect is attributed to the activity of the SWCNTs for initiating the $H_2O$ discharge, thus accelerating the Volmer reaction on the c-$ReS_2$ nanosheets in alkaline conditions.[102] In addition, it is worth noticing that SWCNTs/c-$ReS_2$ displays an HER-activity in acidic media comparable to the one shown by SWCNTs/LPE-$ReS_2$ ($\eta_{10}$ = 0.192 V), which, however, shows a better HER-activity in alkaline media ($\eta_{10}$ = 0.238 V). The morphological and electrochemical characterization of SWCNTs/LPE-$ReS_2$ electrode is reported in the **SI** (see **Figure S10**). Electrochemical impedance spectroscopy measurements do not show any significant differences between SWCNTs/c-$ReS_2$ and SWCNTs/LPE-$ReS_2$ (see Nyquist plot in **Figure S11**). Moreover, we estimated the double-layer capacitance ($C_{dl}$) of the c-$ReS_2$ and LPE-$ReS_2$ films deposited on GC from cyclic voltammetry (CV) measurements (see **SI** for further details, **Figure S12a**). The calculated $C_{dl}$ values, ~0.162 mF $cm^{-2}$ for c-$ReS_2$ film and ~0.110 mF $cm^{-2}$ for LPE-$ReS_2$ film, indicate that the electrochemical accessible surface area of c-$ReS_2$ film is superior to the one of the LPE-$ReS_2$ film. The larger accessible surface area of c-$ReS_2$ (~ 4±1 nm determined by TEM), compared to the LPE-$ReS_2$, can be attributed to the small lateral size of the c-$ReS_2$ nanosheets, as well as their arrangement relatively to the substrate (see **SI**, **Figure S12b-c**).

Beyond the electrocatalytic activity, the durability is another important criterion for the exploitation of an electrocatalyst. **Figure 6c-d** shows the chronoamperometry measurements (current retention *vs.* time) for the SWCNTs/c-$ReS_2$, in acidic and alkaline solutions, respectively. A constant overpotential was applied in order to provide the same starting cathodic current density of 20 mA $cm^{-2}$ for HER. Glassy carbon rod has been used as counter-electrode to avoid Pt dissolution/re-deposition effects altering the HER-activity of working electrode in presence of Pt-based counter electrodes. In both acidic and alkaline media, SWCNTs/c-$ReS_2$ exhibit a nearly stable behavior over 48 h (current retention of 94.3% and 98.0%, respectively). The slight fluctuation of the HER-activity can be tentatively attributed to morphological changes of the electrode film during the HER process. Similar effects have been previously observed in electrocatalysts based on metallic TMDCs, such as $TaS_2$[103] and $NbS_2$[93,104] nanosheets, as well as in other 2D material-based electrodes, including graphene-based electrocatalysts.[105] In fact, the mechanical stresses originated by $H_2$ bubbling can cause a re-orientation/fragmentation of the 2D electrocatalysts, which, consequently, shows higher electrochemically accessible surface area.[103,104,106] Moreover, such mechanical stresses occasionally caused the formation of cracks in the active material films, as observed by SEM image analysis of additional electrodes (**Figures S13**). This led to a significant change of the cathodic current of the electrodes over time in both acidic and alkaline media. For example, in acidic media some electrodes reported an increase of the initial HER-activity over 24 h, achieving a $\eta_0$ = 0.162 V *vs.* RHE (see **Figure S14a**). This behaviour might be attributed to the progressive increase of the $H^+$ accessibility to the HER-active sites of the electrodes.[103,104,106] In alkaline media, the electrodes often displayed an increase of the electrical resistance (*e.g.*, from 4.4 Ω to 45.9 Ω, measured by single frequency EIS), which caused a decrease of 21.5% of the initial current density (see **Figure S14b**). Despite the decrease of the HER-activity of the electrode, the *iR*-corrected polarization curves shows an electrochemical activation of the c-$ReS_2$ nanosheets after 24 h ($\eta_{10}$ reduced from 0.327 V to 0.181 V), similarly to the electrodes tested in acidic media. The use of electrocatalyst binders, such as sulfonated tetrafluoroethylene-based fluoropolymer copolymers (*e.g.*, Nafion), could prospectively "freeze" an optimized electrode morphology of the current electrodes, as well as an increase of the electrode durability under HER-operation.

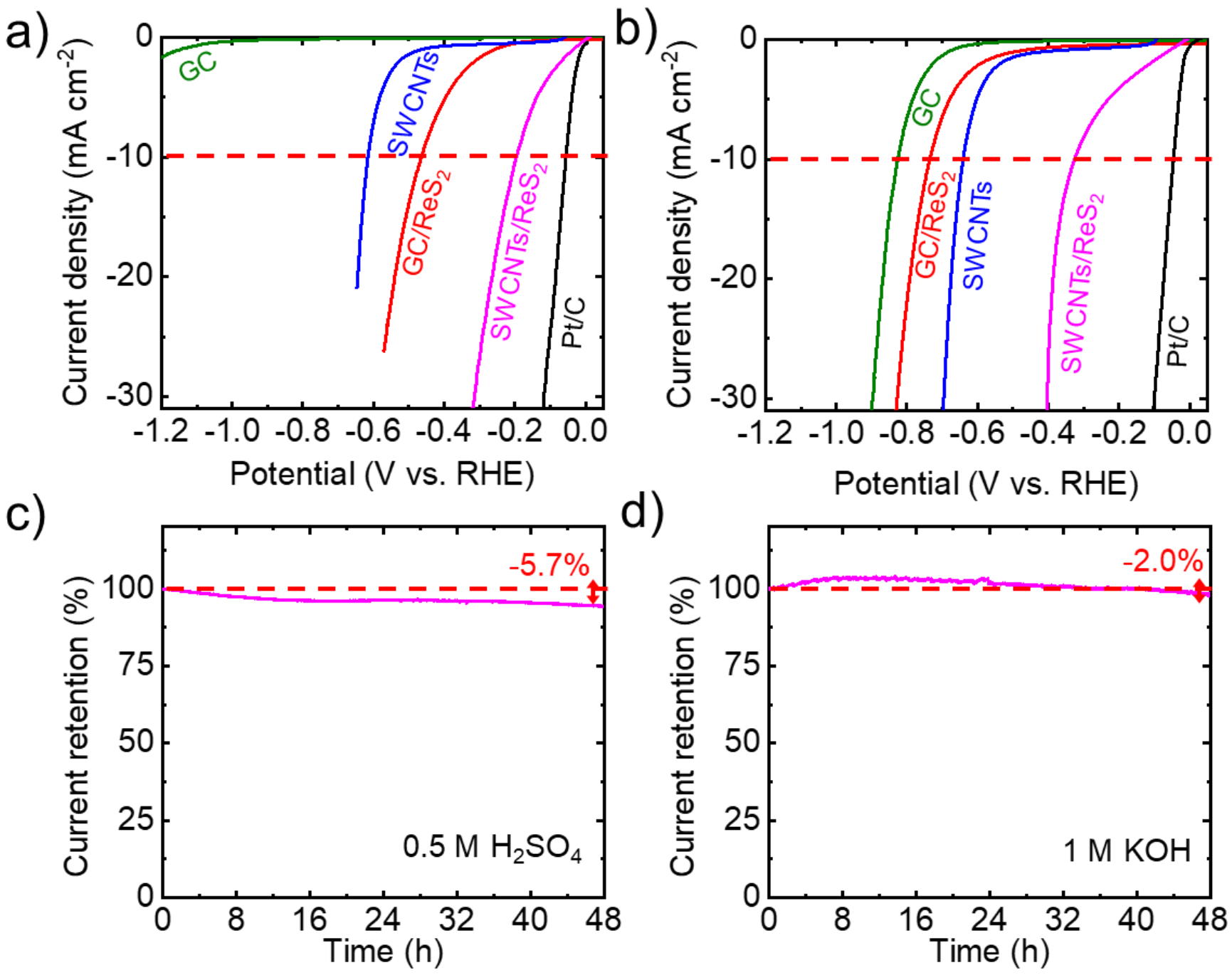

**Figure 6.** (a,b) *iR*-corrected polarization curves measured for GC, SWCNTs, CG/c-$ReS_2$, SWCNTs/c-$ReS_2$ and Pt/C in acidic (0.5 M $H_2SO_4$) and alkaline (1 M KOH) solutions, respectively. (c,d) Chronoamperometry measurements (current retention *vs.* time) for the SWCNTs/c-$ReS_2$ electrode in 0.5 M $H_2SO_4$ and 1 M KOH solutions. A constant overpotential has been applied in order to give an equal starting cathodic current density of 20 mA $cm^{-2}$. The inset panels of c) and d) show the iR-corrected polarization curves after stability tests.

**Table 1** summarizes the $\eta_{10}$ achieved by our electrode in comparison with those reported by different literature results on $ReS_2$-based electrocatalysts, as well as other promising electrocatalysts based on other TMDCs ($MoS_2$,[67,98,107–109] $WS_2$[110] and $MoSe_2$ [61]) and phosphides ($NiCo_2P_x$).[111] In particular, in acidic solution, $\eta_{10}$ of SWCNTs/$ReS_2$ (0.196 V, 0.162 V after 24 h-stability test) approaches the best values obtained by $ReS_2$-based electrocatalysts (0.147 V),[54] which come from a defect-activated monolayer $ReS_2$ grown by CVD. These results indicate that there is still room for improvement in our system by properly optimizing the material surface in future experiments, as it has also been demonstrated for other TMDCs[107,112,113]. Additionally, the HER-performance of our electrocatalysts are comparable or superior to those previously reported for electrocatalysts based on other TMDCs ($\eta_{10}$ values between 0.170 and 0.372 V), concretely, in absence of either mechanical strain or doping effects, including the most studied $MoS_2$,[67,98,107–109] $WS_2$[110] and $MoSe_2$ [61]. The relevant electrocatalytic properties of $ReS_2$ have been recently attributed to the presence of metal-metal bonds (not present for group-$VI_B$ TMDCs),[9,11–13] which are highly HER-active sites in presence of vacancies causing an intrinsic HER-promoting optimization of the electronic charge distribution.[54] Moreover, we reported here also the HER-activity of $ReS_2$-based electrocatalyst in alkaline conditions, providing new insight for the development of pH-universal electrocatalysts to be exploited as cathode materials in current large-scale $H_2$ production technologies,[114],[115],[116] *e.g.*, chloro-alkaline systems[117] or alkaline zero-gap water electrolysis units,[118] and proton exchange membrane (PEM) electrolysis.[119],[120]

**Table 1.** Summary of the $\eta_{10}$ values measured for our electrodes and other $ReS_2$-based electrocatalysts reported in literature in acid and basic conditions. For comparison we also show $\eta_{10}$ values measured for electrocatalysts based on $MoS_2$, $MoSe_2$, $WS_2$ and phosphides.

| Electrocatalyst | $\eta_{10}$ (V) | Electrolyte | Mass loading (mg cm$^{-2}$) | Reference |
|---|---|---|---|---|
| GC/$ReS_2$ | 0.465 | 0.5 M $H_2SO_4$ | ~0.13 | This work |
| | 0.735 | 1 M KOH | | |
| SWCNTs/$ReS_2$ | 0.195 (0.162 after 24 h) | 0.5 M $H_2SO_4$ | ~1.59 | This work |
| | 0.327 (0.181 after 24h) | 1 M KOH | | |
| Chemically exfoliated $ReS_2$ nanosheets | ~0.3$^{a)}$ | 0.5 M $H_2SO_4$ | Not reported | [36] |
| Defects-activated monolayer $ReS_2$ | 0.147 | 0.5 M $H_2SO_4$ | Monolayer | [54] |
| Vertically Oriented Arrays of $ReS_2$ Nanosheets | ~0.3 | 0.5 M $H_2SO_4$ | ~0.67 | [25] |
| Lithiated Vertically Oriented Arrays of $ReS_2$ Nanosheets | ~0.2 | 0.5 M $H_2SO_4$ | ~0.67 (excluding Li) | [25] |
| $ReS_2$ 3D reticulated vitreous carbon foams | 0.336$^{a)}$ | 0.5 M $H_2SO_4$ | Not reported (ReS$_2$ grown onto carbon foam) | [28] |
| One-pot synthetized $ReS_2$ | >0.35 | 0.5 M $H_2SO_4$ | 0.02 | [121] |
| $MoS_2$ | > 0.35 | 0.5 M $H_2SO_4$ | Monolayer | [107] |
| $MoS_2$ with S vacancies | 0.25 | 0.5 M $H_2SO_4$ | | |
| Strained $MoS_2$ with S vacancies | 0.17 | 0.5 M $H_2SO_4$ | | |
| Co-doped $MoS_2$ | 0.159 | 0.5 M $H_2SO_4$ | 0.5 | [112] |
| Reduced graphene oxide:$MoS_2$ hybrid on GC | ~0.15 | 0.5 M $H_2SO_4$ | 0.285 | [113] |
| Reduced graphene oxide:$MoS_2$ hybrid on carbon fibers | ~0.15 | 0.5 M $H_2SO_4$ | 1 | |
| $MoS_2$ nanosheets | > 0.3 | 0.5 M $H_2SO_4$ | 0.05 | [67] |
| Metallic (1T) $MoS_2$ nanosheets | ~0.20 | 0.5 M $H_2SO_4$ | 0.05 | |
| Liquid-phase exfoliated $MoS_2$ | 0.372 | 0.5 M $H_2SO_4$ | 0.5 | [98] |
| Graphene/$MoS_2$ | 0.175 | 0.5 M $H_2SO_4$ | 0.5 | |
| Chemically exfoliated metallic (1T)-$MoS_2$ | 0.235 | 0.5 M $H_2SO_4$ | 0.5 | |
| CVD-grown $MoS_2$ monolayer on GC | 0.342 | 0.5 M $H_2SO_4$ | Not reported | [108] |
| Solvothermal produced $MoS_2$ on GC | 0.252 | 0.5 M $H_2SO_4$ | ~0.5 | [109] |
| Liquid-phase exfoliated $MoSe_2$ | 0.37 | 0.5 M $H_2SO_4$ | 2 | [61] |
| SWCNTs/$MoSe_2$ | 0.17 | 0.5 M $H_2SO_4$ | 2 | [61] |
| $WS_2$ nanosheets | > 150 | 0.5 M $H_2SO_4$ | 0.35 | [110] |
| $NiCo_2P_x$ Nanowires | 0.104 | 0.5 M $H_2SO_4$ | 5.9 | [111] |

a) Data not iR-corrected

## 3. Conclusions

We report a bottom-up approach for the synthesis of colloidal $ReS_2$ (c-$ReS_2$) nanosheets that relies on low fabrication temperatures (below 360 °C) and process times inferior to 4 h. We use elemental sulfur as source, which is more accessible and environmentally friendly than other precursor materials currently used in the colloidal synthesis of different TMDCs. Drop-cast c-$ReS_2$-based films are tested as gas sensors for a variety of agents, achieving highly competitive performance after annealing or ligand exchange in comparison with devices built with CVD-$ReS_2$ or -$MoS_2$. Furthermore, the colloidal $ReS_2$ nanosheets were tested as HER-electrocatalysts operating both in alkaline and acidic media. In particular, electrodes made by c-$ReS_2$ films deposited on single-walled carbon nanotubes (SWCNT) (SWCNTs/c-$ReS_2$) exhibit overpotentials at a cathodic current density of 10 mA $cm^{-2}$ ($\eta_{10}$) of 0.196 V in 0.5 M in $H_2SO_4$ and 0.299 V in 1 M KOH. In acidic media, the $\eta_{10}$ of SWCNTs/c-$ReS_2$ approaches the best values obtained by CVD-grown defect-activated $ReS_2$-based electrocatalysts. In comparison with non-optimized electrocatalysts based on other group-$VI_B$ TMDCs, the $ReS_2$-based electrocatalysts here developed exhibit similar or even superior HER-activity. The c-$ReS_2$ nanosheets can therefore present a promising solution for the fabrication of gas sensors and HER electrocatalysts, with still considerable room for improvement by exploiting device designs and finely tuned surface modifications.

## 4. Experimental Section

*4.1. Materials.* Rhenium (V) chloride ($ReCl_5$, 99.9%, Alfa Aesar); sulfur powder (99%, Strem Chemicals); oleic acid (OA, 90%, Sigma Aldrich®, degassed at 100°C 2h); oleylamine (OlAm, 70%, Sigma Aldrich®); 1-octadecene (ODE, 90%, Sigma Aldrich®, degassed at 150°C 3h); acetone (≥99.5%, Sigma Aldrich®); ethanol (EtOH, (≥99.8, without additive, Sigma Aldrich®); methanol (MeOH, ≥99.8%, Sigma Aldrich®); 2-propanol (IsOH, ≥99.8%, Sigma Aldrich®); 3-mercaptopropionic acid (MPA, ≥99%, Sigma Aldrich®); 4-aminobenzoic acid (ABA, ≥99%, Sigma Aldrich®); 1,4-benzenedithiol (BDT, 97%, Alfa Aesar). Au pellets (99.999%) were purchased from Kurt J. Lesker.

*4.2. Colloidal synthesis of $ReS_2$ nanosheets.* The general procedure to obtain c-$ReS_2$ consists in the use of two separated precursor solutions: one in the round-bottom flask and the second one added by syringe pump method. For the reaction, we used a metal:chalcogenide molar ratio of 1:6. The dispersion that contains the Re precursor ($ReCl_5$ 67 mg) is prepared by sonication in a bath at 60°C (1h) with 500 µL OA + 2 mL ODE. This will be added by syringe pump (rate 2 mL $h^{-1}$) in a hot medium under Ar flow (350°C) of S-OlAm. The synthesis medium itself was prepared by mixing 35 mg S with 7 mL OlAm and degassing under Ar during 2h at 250°C in a Schlenk line until a clear orange-brownish solution is achieved. The reaction temperature was set to 350°C. The suspension as obtained from synthesis was purified twice (in air) by adding toluene, acetone and isopropanol (15:10:5 mL), followed by centrifugation (2599 *g*, Sigma 3-16P centrifuge, rotor 19776). The sheets were finally dispersed in toluene using a vortex.

*4.3. TEM.* The morphology of the synthesized c-$ReX_2$ nanosheets and LPE-$ReX_2$ (X = S or Se) flakes was evaluated by transmission electron microscopy in a JEOL JEM-1011 microscope (W filament), operated at 100 kV. Samples were drop-cast on carbon-coated copper grids. Selected area electron diffraction patterns and overview TEM images were also acquired with a JEOL JEM-1400Plus TEM, with a thermionic source ($LaB_6$ crystal), operated at 120 kV. Selected area electron diffraction pattern processing (azimuthal integration, background subtraction) was done using the PASAD plugin for Digital Micrograph.[122] High-angle annular dark-field scanning TEM and bright-field TEM imaging at higher magnification of the c-$ReX_2$ nanosheets (X = S or Se) samples were carried out by using a FEI Tecnai $G^2$ F20, with Schottky emitter, operated at 200 kV. Statistics for the estimation of the lateral size and thickness were carried out considering 20 individual domains in the nanosheets for each sample.

*4.4. AFM.* Atomic force microscopy (AFM) images were acquired by using a Nanowizard III (JPK Instruments, Germany) mounted on an Axio Observer D1 (Carl Zeiss, Germany) inverted optical microscope. The samples were prepared by drop-casting c$ReS_2$ flakes dispersion onto a freshly cleaved mica substrate (G250-1, Agar Scientific Ltd., Essex, U.K.) and the AFM measurements were carried out by using PPP-NCHR cantilevers (Nanosensors, USA) with a nominal tip diameter of 10 nm. Intermittent contact mode AFM images of 5×5 $\mu m^2$ and 1.5×1.5 $\mu m^2$ were collected with 1024 data points per line and the working set point is kept above 70% of the free oscillation amplitude. The scan rate for the acquisition of images was 0.9 Hz. Height profiles were processed by using the JPK Data Processing software (JPK Instruments, Germany) and the data were analyzed with OriginPro 9.1 software.
*4.5. XPS.* X-ray photoelectron spectroscopy was carried out in a Kratos Axis UltraDLD spectrometer using a monochromatic Al K a source (15 kV, 20 mA). High-resolution scans were performed at a constant pass energy of 10 eV and steps of 0.1 eV. The photo-electrons were detected at a take-off angle φ = 0° with respect to the surface normal. The pressure in the analysis chamber was kept below $7 \times 10^{-9}$ Torr. The binding energy scale was internally referenced to the Au $4f^{7/2}$ peak at 84 eV. The spectra were analyzed using the CasaXPS software (version 2.3.16).
*4.6. Raman spectroscopy characterization.* Raman measurements were performed in a Renishaw inVia micro-Raman microscope equipped with a 50× (0.75 N.A.) objective with excitation wavelengths of 532 and 785 nm and an incident power ≤ 1 mW to avoid heating and damage of the samples. The samples were prepared by drop casting the diluted materials dispersion onto a Si wafer covered with 300 nm thermally grown $SiO_2$ (LDB Technologies Ltd.). For each sample, at least 50 spectra were collected.
*4.7. Fabrication of the gas sensors.* A ligand exchange procedure in solution was carried out with the as-prepared OA/OlAm-capped $ReS_2$ nanosheets dispersed in toluene. The different ligands (MPA, ATP and BDT) were dissolved in MeOH at a concentration of 1 mM. The c-$ReS_2$ dispersion in toluene (50 mg $mL^{-1}$) and the ligand solution were mixed in a volume ratio of 1:1 and stirred 2 min in a vortex, following by centrifugation (5000 rpm) and removal of the supernatant. Then, procedure is repeated once more, with a subsequent purification step with toluene: MeOH twice and centrifugation. Finally, the ligand exchanged $ReS_2$ nanosheets were dispersed in isopropanol at a concentration of 50 mg $mL^{-1}$, and immediately used for the preparation of the films on cleaned glass slides (ca. 1.5×1.5 $cm^2$) by drop-casting (100 µL). Prior to the deposition, the glass substrates were cleaned in an ultrasonic bath (8 min each step), first with acetone, followed by isopropanol, dried with a $N_2$ flow, and plasma cleaning step under $N_2$ (100W, 2 min). For the Au contacts deposition, we used a shadow mask with square holes of 1×1 $mm^2$ area, separated by 100 µm. The Au film with a thickness of 80 nm was evaporated in a Kenosistec® e-beam evaporator at a deposition rate of 0.3 Å $s^{-1}$ and base pressure of about 1.0 $10^{-6}$ mbar. The thickness of the different films was measured with a Veeco Dektak 150 profilometer.
*4.8. Electrical characterization of the gas sensors.* We carried out the electrical tests in a chamber (see **Figure 3a**) equipped with two inlets, one for the $N_2$ and the other for the target gases. Side ports allows the access of micromanipulators connected to a Keithley 2612 source-meter. The chamber (~643 $cm^3$) has a transparent top lid, such that positioning of the tips is possible by eye. A $N_2$ line, at a pressure slightly above ambient pressure (1.2 bar, flow of ~ 667 $cm^3$ $s^{-1}$ determined with a flowmeter (Yokogawa®)), provided the inert atmosphere used as reference and allows to fill and purge the chamber in ~1 s. A glass bubbler was connected and used for containing the water (MilliQ®), ammonia (2% vol aq. solution), EtOH and acetone while bubbling with the $N_2$ flow used as carrier to create the saturated atmosphere. To ensure uniform evaporation of the volatile gases, the bubbler was kept in a warm water bath (60°C). For $CO_2$ and compressed dry air, reservoirs connected to the experimental chamber were used. The experiments were performed according to the following procedure: current was measured as a function of time, first under $N_2$ flux, then the test gas was injected in the chamber for 10s, then the flux was switched back to $N_2$ until recovery of the base current. The procedure was repeated multiple times for each device to ensure reproducibility.

*4.9. Fabrication of the electrocatalyst electrodes.* The solvent was removed from the as-prepared (c-$ReS_2$) sample by bubbling with nitrogen. The obtained powder was annealed in a tubular furnace (PSC 12/--/600H, Lenton, UK, 25 mm inner tube diameter, Ar 100sccm) at 500 °C for 8 h. Then, the powder was dispersed in NMP at a concentration of 1 mg $mL^{-1}$ and sonicated for 7 h. We prepared the GC/c-$ReS_2$ electrodes by drop-casting the c-$ReS_2$ NMP dispersion (0.2 mg) on GC substrates (Sigma Aldrich®) with geometrical area of 1×1.5 $cm^2$ (c-$ReS_2$ mass loading ~0.13 mg $cm^{-2}$). Electrodes of Pt/C were produced as benchmark for HER by depositing the Pt/C dispersion onto GC substrates. The Pt/C dispersion was produced by dissolving 4 mg of Pt/C (5 wt.% Pt loading, Sigma Aldrich®) and 80 µL of Nafion solution (5 wt.%, Sigma Aldrich®) in 1 mL of 1:4 v/v ethanol/water. The Pt/C mass loading of the electrodes was 0.262 mg $cm^{-2}$, in agreement with protocols developed previously.[102,105]. The SWCNTs/c-$ReS_2$ electrodes were fabricated by sequential vacuum filtrations of 5 mg of SWCNTs (>90%, Cheap Tubes, sonic-tip de-bundled in NMP at a 0.4 mg $mL^{-1}$ concentration[61,63]) and 5 mg of c-$ReS_2$ on nylon filter membranes (0.2 µm pore size, 25 mm diameter from Sigma Aldrich®). The as-produced films (geometrical area = 3.5 $cm^2$, and c-$ReS_2$ mass loading ~1.59 mg $cm^{-2}$) spontaneously peels off the nylon membrane during the drying, resulting in self-standing electrodes ready to be used. The SWCNTs/LPE-$ReS_2$ electrodes were produced using the same protocols described for SWCNTs/c-$ReS_2$, except for the use of the as-produced LPE-$ReS_2$ dispersion in isopropanol instead of c-$ReS_2$ one. Additional electrodes were produced by depositing the c-$ReS_2$ and LPE-$ReS_2$ nanosheet dispersions onto GC by drop casting method (catalyst mass loading = 0.2 mg $cm^{-2}$). All the electrodes were dried overnight at room temperature before their electrochemical characterization.

*4.10. Electrochemical measurements of the electrodes.* Measurements were carried out at room temperature in a flat-bottom quartz cell under a three-electrode configuration using an Ivium® CompactStat potentiostat/galvanostat station controlled *via* IviumSoft®. A Pt wire or a GC rod were used as the counter electrode and a KCl-saturated Ag/AgCl was used as the reference electrode. As aqueous medium, 200 mL of two aqueous solutions were used: 0.5 M $H_2SO_4$ (99.999%, Sigma-Aldrich) and 1M KOH (90%, Sigma Aldrich). MilliQ® water was used to prepare the solutions. Oxygen was purged from electrolyte solutions by flowing $N_2$ throughout the aqueous medium using a porous frit for 30 min before starting the measurements. A constant $N_2$ flow was kept afterward for the whole duration of the experiments to avoid re-dissolution of $O_2$ in the electrolyte. The potential difference between the working electrode and the Ag/AgCl reference electrode was converted to the RHE scale using the Nernst equation: $E_{RHE} = E_{Ag/AgCl} + 0.059pH + E^0_{Ag/AgCl}$ , where $E_{RHE}$ is the converted potential *vs.* RHE, $E_{Ag/AgCl}$ is the potential experimentally measured against the Ag/AgCl reference electrode, and $E^0_{Ag/AgCl}$ is the standard potential of Ag/AgCl at 25 °C (0.1976 V). The LSV curves were acquired at the scan rate of 5 mV $s^{-1}$. Polarization curves were *iR*-corrected, in which *i* is the measured working electrode current and the *R* is the series resistance that arises from the working electrode substrate and electrolyte resistances. *R* was extrapolated by the real part of the impedance (Re[Z]) measured by single frequency EIS at open-circuit potential and at the frequency of 100 kHz. Electrochemical impedance spectra of the SWCNTs/c-$ReS_2$ and SWCNTs/LPE-$ReS_2$ were acquired at open circuit potential at frequencies between 0.1 Hz and 200 kHz. Stability tests were carried out by chronoamperometry measurements (*j*–*t* curves) by measuring the current in the potentiostatic mode at a fixed overpotential in order to provide the same starting cathodic current density of 20 mA $cm^{-2}$ for HER. The $C_{dl}$ of the c-$ReS_2$ and LPE-$ReS_2$ films deposited onto GC were estimated by CV measurements in a non-Faradaic region of potential (*i.e.*, potential between 0.30 and 0.45 V *vs.* Ag/AgCl) at various potential scan rates (from 20 to 800 mV $s^{-1}$) in 0.5 M $H_2SO_4$.

*4.11. SEM/EDX characterization.* We performed SEM/EDX measurements on the sensor devices and electrode samples using a Helios Nanolab 600 (FEI Company) microscope. SEM measurements were performed at 5 kV and 0.2 nA. For the cross-section images, we carefully cut the electrodes with a scalpel and measured in 90° tilted sample holder.

## Acknowledgements

This project has received funding from the European Union's Horizon 2020 research and innovation programme under grant agreement no. 785219 (GrapheneCore2), and the European Research Council (grant agreement no. 714876 PHOCONA). The authors thank Dr. A. Toma for the access to the IIT clean room facilities for the SEM measurements and evaporation procedures, Dr. F. de Angelis for the access to the Raman lab and the Materials Characterization and Electron Microscopy facilities for the access to the XPS equipment and TEM, respectively. M. Leoncini is acknowledged for the support in the fabrication of the gas chamber.

## References

[1] R. Lv, J. A. Robinson, R. E. Schaak, D. Sun, Y. Sun, T. E. Mallouk, M. Terrones, *Acc. Chem. Res.* **2015**, *48*, 56.

[2] Z. Lin, A. McCreary, N. Briggs, S. Subramanian, K. Zhang, Y. Sun, X. Li, N. J. Borys, H. Yuan, S. K. Fullerton-Shirey, A. Chernikov, H. Zhao, S. McDonnell, A. M. Lindenberg, K. Xiao, B. J. LeRoy, M. Drndić, J. C. M. Hwang, J. Park, M. Chhowalla, R. E. Schaak, A. Javey, M. C. Hersam, J. Robinson, M. Terrones, *2D Mater.* **2016**, *3*, 042001.

[3] J. H. Han, M. Kwak, Y. Kim, J. Cheon, *Chem. Rev.* **2018**, *118*, 6151.

[4] W. S. Yun, S. W. Han, S. C. Hong, I. G. Kim, J. D. Lee, *Phys. Rev. B* **2012**, *85*, 033305.

[5] P. Tonndorf, R. Schmidt, P. Böttger, X. Zhang, J. Börner, A. Liebig, M. Albrecht, C. Kloc, O. Gordan, D. R. T. Zahn, S. Michaelis de Vasconcellos, R. Bratschitsch, *Opt. Express* **2013**, *21*, 4908.

[6] X. Chia, A. Y. S. Eng, A. Ambrosi, S. M. Tan, M. Pumera, *Chem. Rev.* **2015**, *115*, 11941.

[7] X. Liu, T. Galfsky, Z. Sun, F. Xia, E. Lin, Y.-H. Lee, S. Kéna-Cohen, V. M. Menon, *Nat. Photonics* **2014**, *9*, 30.

[8] A. A. Tedstone, D. J. Lewis, P. O'Brien, *Chem. Mater.* **2016**, *28*, 1965.

[9] M. Rahman, K. Davey, S.-Z. Qiao, *Adv. Funct. Mater.* **2017**, *27*, 1606129.

[10] M. Hafeez, L. Gan, A. Saleem Bhatti, T. Zhai, *Mater. Chem. Front.* **2017**, *1*, 1917.

[11] D. A. Chenet, O. B. Aslan, P. Y. Huang, C. Fan, A. M. van der Zande, T. F. Heinz, J. C. Hone, *Nano Lett.* **2015**, *15*, 5667.

[12] H. H. Murray, S. P. Kelty, R. R. Chianelli, C. S. Day, *Inorg. Chem.* **1994**, *33*, 4418.

[13] S. P. Kelty, A. F. Ruppert, R. R. Chianelli, J. Ren, M.-H. Whangbo, *J. Am. Chem. Soc.* **1994**, *116*, 7857.

[14] K. Momma, F. Izumi, *J. Appl. Crystallogr.* **2011**, *44*, 1272.

[15] S. Tongay, H. Sahin, C. Ko, A. Luce, W. Fan, K. Liu, J. Zhou, Y.-S. Huang, C.-H. Ho, J. Yan, D. F. Ogletree, S. Aloni, J. Ji, S. Li, J. Li, F. M. Peeters, J. Wu, *Nat. Commun.* **2014**, *5*, 3252.

[16] H. Jang, C. R. Ryder, J. D. Wood, M. C. Hersam, D. G. Cahill, *Adv. Mater.* **2017**, *29*, 1700650.

[17] O. B. Aslan, D. A. Chenet, A. M. van der Zande, J. C. Hone, T. F. Heinz, *ACS Photonics* **2016**, *3*, 96.
[18] S.-H. Jo, H.-Y. Park, D.-H. Kang, J. Shim, J. Jeon, S. Choi, M. Kim, Y. Park, J. Lee, Y. J. Song, S. Lee, J.-H. Park, *Adv. Mater.* **2016**, *28*, 6711.
[19] M. Hafeez, L. Gan, H. Li, Y. Ma, T. Zhai, *Adv. Funct. Mater.* **2016**, *26*, 4551.
[20] B. Jariwala, D. Voiry, A. Jindal, B. A. Chalke, R. Bapat, A. Thamizhavel, M. Chhowalla, M. Deshmukh, A. Bhattacharya, *Chem. Mater.* **2016**, *28*, 3352.
[21] G. Nazir, M. A. Rehman, M. F. Khan, G. Dastgeer, S. Aftab, A. M. Afzal, Y. Seo, J. Eom, *ACS Appl. Mater. Interfaces* **2018**, *10*, 32501.
[22] S. Yang, J. Kang, Q. Yue, J. M. D. Coey, C. Jiang, *Adv. Mater. Interfaces* **2016**, *3*, 1500707.
[23] A. Yang, J. Gao, B. Li, J. Tan, Y. Xiang, T. Gupta, L. Li, S. Suresh, J. C. Idrobo, T.-M. Lu, M. Rong, N. Koratkar, *2D Mater.* **2016**, *3*, 045012.
[24] Q. Zhang, S. Tan, R. G. Mendes, Z. Sun, Y. Chen, X. Kong, Y. Xue, M. H. Rümmeli, X. Wu, S. Chen, L. Fu, *Adv. Mater.* **2016**, *28*, 2616.
[25] J. Gao, L. Li, J. Tan, H. Sun, B. Li, J. C. Idrobo, C. V. Singh, T.-M. Lu, N. Koratkar, *Nano Lett.* **2016**, *16*, 3780.
[26] Q. Li, Y. Xu, Z. Yao, J. Kang, X. Liu, C. Wolverton, M. C. Hersam, J. Wu, V. P. Dravid, *ACS Nano* **2018**, *12*, 7875.
[27] A.-J. Cho, S. D. Namgung, H. Kim, J.-Y. Kwon, *APL Mater.* **2017**, *5*, 076101.
[28] L. Wang, Z. Sofer, J. Luxa, D. Sedmidubský, A. Ambrosi, M. Pumera, *Electrochem. Commun.* **2016**, *63*, 39.
[29] Q. Zhang, W. Wang, J. Zhang, X. Zhu, Q. Zhang, Y. Zhang, Z. Ren, S. Song, J. Wang, Z. Ying, R. Wang, X. Qiu, T. Peng, L. Fu, *Adv. Mater.* **2018**, *30*, 1707123.
[30] "Live PGM Prices," can be found under https://www.metalsdaily.com/live-prices/pgms/, **n.d.**
[31] "Engelhard Industrial Bullion (EIB) Prices," can be found under https://apps.catalysts.basf.com/apps/eibprices/mp/, **n.d.**
[32] K. Keyshar, Y. Gong, G. Ye, G. Brunetto, W. Zhou, D. P. Cole, K. Hackenberg, Y. He, L. Machado, M. Kabbani, A. H. C. Hart, B. Li, D. S. Galvao, A. George, R. Vajtai, C. S. Tiwary, P. M. Ajayan, *Adv. Mater.* **2015**, *27*, 4640.
[33] N. Al-Dulaimi, D. J. Lewis, X. L. Zhong, M. Azad Malik, P. O'Brien, *J Mater Chem C* **2016**, *4*, 2312.
[34] Y. Kim, B. Kang, Y. Choi, J. H. Cho, C. Lee, *2D Mater.* **2017**, *4*, 025057.
[35] F. Cui, C. Wang, X. Li, G. Wang, K. Liu, Z. Yang, Q. Feng, X. Liang, Z. Zhang, S. Liu, Z. Lei, Z. Liu, H. Xu, J. Zhang, *Adv. Mater.* **2016**, *28*, 5019.
[36] T. Fujita, Y. Ito, Y. Tan, H. Yamaguchi, D. Hojo, A. Hirata, D. Voiry, M. Chhowalla, M. Chen, *Nanoscale* **2014**, *6*, 12458.
[37] N. Al-Dulaimi, E. A. Lewis, D. J. Lewis, S. K. Howell, S. J. Haigh, P. O'Brien, *Chem. Commun.* **2016**, *52*, 7878.
[38] J. Kang, V. K. Sangwan, J. D. Wood, X. Liu, I. Balla, D. Lam, M. C. Hersam, *Nano Lett.* **2016**, *16*, 7216.
[39] H. Weller, J. Niehaus, *Apparatus and Method for the Manufacture of Nanoparticles*, **2009**, GB2457314A.
[40] H. Weller, J. Niehaus, *Reactor for the Manufacture of Nanoparticles*, **2009**, WO2009101091A1.
[41] H. Weller, J. Niehaus, *Reactor for the Manufacture of Nanoparticles*, **2015**, US9084979B2.
[42] D. Son, S. I. Chae, M. Kim, M. K. Choi, J. Yang, K. Park, V. S. Kale, J. H. Koo, C. Choi, M. Lee, J. H. Kim, T. Hyeon, D.-H. Kim, *Adv. Mater.* **2016**, *28*, 9326.
[43] X. Li, A. Tang, J. Li, L. Guan, G. Dong, F. Teng, *Nanoscale Res. Lett.* **2016**, *11*, 171.
[44] M. Alam Khan, Y.-M. Kang, *J. Energy Storage* **2016**, *7*, 252.
[45] Y. Sun, F. Alimohammadi, D. Zhang, G. Guo, *Nano Lett.* **2017**, *17*, 1963.
[46] W. Jung, S. Lee, D. Yoo, S. Jeong, P. Miró, A. Kuc, T. Heine, J. Cheon, *J. Am. Chem. Soc.* **2015**, *137*, 7266.
[47] D. Sun, S. Feng, M. Terrones, R. E. Schaak, *Chem. Mater.* **2015**, *27*, 3167.

[48] Y. Sun, Y. Wang, D. Sun, B. R. Carvalho, C. G. Read, C. Lee, Z. Lin, K. Fujisawa, J. A. Robinson, V. H. Crespi, M. Terrones, R. E. Schaak, *Angew. Chem. Int. Ed.* **2016**, *55*, 2830.
[49] B. Mahler, V. Hoepfner, K. Liao, G. A. Ozin, *J. Am. Chem. Soc.* **2014**, *136*, 14121.
[50] D. Yoo, M. Kim, S. Jeong, J. Han, J. Cheon, *J. Am. Chem. Soc.* **2014**, *136*, 14670.
[51] S. Yang, C. Jiang, S. Wei, *Appl. Phys. Rev.* **2017**, *4*, 021304.
[52] Z. Yin, H. Li, H. Li, L. Jiang, Y. Shi, Y. Sun, G. Lu, Q. Zhang, X. Chen, H. Zhang, *ACS Nano* **2012**, *6*, 74.
[53] F. Xia, T. Mueller, Y. Lin, A. Valdes-Garcia, P. Avouris, *Nat. Nanotechnol.* **2009**, *4*, 839.
[54] Y. Zhou, E. Song, J. Zhou, J. Lin, R. Ma, Y. Wang, W. Qiu, R. Shen, K. Suenaga, Q. Liu, J. Wang, Z. Liu, J. Liu, *ACS Nano* **2018**, *12*, 4486.
[55] S. E. Hosseini, M. A. Wahid, *Renew. Sustain. Energy Rev.* **2016**, *57*, 850.
[56] J. A. Turner, *Science* **2004**, *305*, 972.
[57] I. Roger, M. A. Shipman, M. D. Symes, *Nat. Rev. Chem.* **2017**, *1*, 0003.
[58] S. Bellani, L. Najafi, A. Capasso, A. E. Del Rio Castillo, M. R. Antognazza, F. Bonaccorso, *J. Mater. Chem. A* **2017**, *5*, 4384.
[59] J. Yang, H. S. Shin, *J Mater Chem A* **2014**, *2*, 5979.
[60] J. D. Benck, T. R. Hellstern, J. Kibsgaard, P. Chakthranont, T. F. Jaramillo, *ACS Catal.* **2014**, *4*, 3957.
[61] L. Najafi, S. Bellani, R. Oropesa-Nuñez, A. Ansaldo, M. Prato, A. E. Del Rio Castillo, F. Bonaccorso, *Adv. Energy Mater.* **2018**, *8*, 1703212.
[62] W. Xiao, P. Liu, J. Zhang, W. Song, Y. P. Feng, D. Gao, J. Ding, *Adv. Energy Mater.* **2017**, *7*, 1602086.
[63] L. Najafi, S. Bellani, R. Oropesa-Nuñez, A. Ansaldo, M. Prato, A. E. Del Rio Castillo, F. Bonaccorso, *Adv. Energy Mater.* **2018**, *8*, 1801764.
[64] J. Zhang, T. Wang, P. Liu, S. Liu, R. Dong, X. Zhuang, M. Chen, X. Feng, *Energy Environ. Sci.* **2016**, *9*, 2789.
[65] M. A. R. Anjum, H. Y. Jeong, M. H. Lee, H. S. Shin, J. S. Lee, *Adv. Mater.* **2018**, *30*, 1707105.
[66] J. Kibsgaard, Z. Chen, B. N. Reinecke, T. F. Jaramillo, *Nat. Mater.* **2012**, *11*, 963.
[67] D. Voiry, M. Salehi, R. Silva, T. Fujita, M. Chen, T. Asefa, V. B. Shenoy, G. Eda, M. Chhowalla, *Nano Lett.* **2013**, *13*, 6222.
[68] B. Hinnemann, P. G. Moses, J. Bonde, K. P. Jørgensen, J. H. Nielsen, S. Horch, I. Chorkendorff, J. K. Nørskov, *J. Am. Chem. Soc.* **2005**, *127*, 5308.
[69] Y. Yu, S.-Y. Huang, Y. Li, S. N. Steinmann, W. Yang, L. Cao, *Nano Lett.* **2014**, *14*, 553.
[70] Y. Jiao, A. M. Hafez, D. Cao, A. Mukhopadhyay, Y. Ma, H. Zhu, *Small* **2018**, *14*, 1800640.
[71] A. Ambrosi, Z. Sofer, M. Pumera, *Chem. Commun.* **2015**, *51*, 8450.
[72] T. F. Jaramillo, K. P. Jorgensen, J. Bonde, J. H. Nielsen, S. Horch, I. Chorkendorff, *Science* **2007**, *317*, 100.
[73] D. Yoo, M. Kim, S. Jeong, J. Han, J. Cheon, *J. Am. Chem. Soc.* **2014**, *136*, 14670.
[74] R. F. Bacon, E. S. Boe, *Ind. Eng. Chem.* **1945**, *37*, 469.
[75] S. Tongay, H. Sahin, C. Ko, A. Luce, W. Fan, K. Liu, J. Zhou, Y.-S. Huang, C.-H. Ho, J. Yan, D. F. Ogletree, S. Aloni, J. Ji, S. Li, J. Li, F. M. Peeters, J. Wu, *Nat. Commun.* **2014**, *5*, 3252.
[76] D. Chen, Y. Gao, Y. Chen, Y. Ren, X. Peng, *Nano Lett.* **2015**, *15*, 4477.
[77] S. A. Dalmatova, A. D. Fedorenko, L. N. Mazalov, I. P. Asanov, A. Yu. Ledneva, M. S. Tarasenko, A. N. Enyashin, V. I. Zaikovskii, V. E. Fedorov, *Nanoscale* **2018**, *10*, 10232.
[78] C. Ronning, H. Feldermann, R. Merk, H. Hofsäss, P. Reinke, J.-U. Thiele, *Phys. Rev. B* **1998**, *58*, 2207.
[79] C. H. Ho, Y. S. Huang, K. K. Tiong, P. C. Liao, *Phys. Rev. B* **1998**, *58*, 16130.
[80] J. Tauc, *Mater. Res. Bull.* **1968**, *3*, 37.
[81] J. I. Pankove, D. A. Kiewit, *J. Electrochem. Soc.* **1972**, *119*, 156C.
[82] H. Liu, B. Xu, J.-M. Liu, J. Yin, F. Miao, C.-G. Duan, X. G. Wan, *Phys. Chem. Chem. Phys.* **2016**, *18*, 14222.

[83] N. R. Pradhan, A. McCreary, D. Rhodes, Z. Lu, S. Feng, E. Manousakis, D. Smirnov, R. Namburu, M. Dubey, A. R. Hight Walker, H. Terrones, M. Terrones, V. Dobrosavljevic, L. Balicas, *Nano Lett.* **2015**, *15*, 8377.
[84] A. McCreary, J. R. Simpson, Y. Wang, D. Rhodes, K. Fujisawa, L. Balicas, M. Dubey, V. H. Crespi, M. Terrones, A. R. Hight Walker, *Nano Lett.* **2017**, *17*, 5897.
[85] G. Gouadec, P. Colomban, *Prog. Cryst. Growth Charact. Mater.* **2007**, *53*, 1.
[86] W. Yang, L. Gan, H. Li, T. Zhai, *Inorg. Chem. Front.* **2016**, *3*, 433.
[87] X. Tang, A. Du, L. Kou, *Wiley Interdiscip. Rev. Comput. Mol. Sci.* **2018**, *8*, e1361.
[88] P. R. Brown, D. Kim, R. R. Lunt, N. Zhao, M. G. Bawendi, J. C. Grossman, V. Bulović, *ACS Nano* **2014**, *8*, 5863.
[89] G. H. Carey, A. L. Abdelhady, Z. Ning, S. M. Thon, O. M. Bakr, E. H. Sargent, *Chem. Rev.* **2015**, *115*, 12732.
[90] D. J. Late, Y.-K. Huang, B. Liu, J. Acharya, S. N. Shirodkar, J. Luo, A. Yan, D. Charles, U. V. Waghmare, V. P. Dravid, C. N. R. Rao, *ACS Nano* **2013**, *7*, 4879.
[91] S.-Y. Cho, S. J. Kim, Y. Lee, J.-S. Kim, W.-B. Jung, H.-W. Yoo, J. Kim, H.-T. Jung, *ACS Nano* **2015**, *9*, 9314.
[92] E. Petroni, E. Lago, S. Bellani, D. W. Boukhvalov, A. Politano, B. Gürbulak, S. Duman, M. Prato, S. Gentiluomo, R. Oropesa-Nuñez, J.-K. Panda, P. S. Toth, A. E. Del Rio Castillo, V. Pellegrini, F. Bonaccorso, *Small* **2018**, *14*, 1800749.
[93] L. Najafi, S. Bellani, R. Oropesa-Nuñez, B. Martin-Garcia, M. Prato, V. Mazanek, D. Debellis, S. Laucello, R. Brescia, Z. Sofer, F. Bonaccorso, *J Mater Chem A* **2019**, *just accepted*, DOI 10.1039/c9ta07210a.
[94] K. A. Mauritz, R. B. Moore, *Chem. Rev.* **2004**, *104*, 4535.
[95] P. S. Khadke, U. Krewer, *J. Phys. Chem. C* **2014**, *118*, 11215.
[96] K. Shinozaki, B. S. Pivovar, S. S. Kocha, *ECS Trans.* **2013**, *58*, 15.
[97] H.-C. Tu, W.-L. Wang, C.-C. Wan, Y.-Y. Wang, *J. Phys. Chem. B* **2006**, *110*, 15988.
[98] L. Najafi, S. Bellani, B. Martín-García, R. Oropesa-Nuñez, A. E. Del Rio Castillo, M. Prato, I. Moreels, F. Bonaccorso, *Chem. Mater.* **2017**, *29*, 5782.
[99] Q. Liu, Q. Fang, W. Chu, Y. Wan, X. Li, W. Xu, M. Habib, S. Tao, Y. Zhou, D. Liu, T. Xiang, A. Khalil, X. Wu, M. Chhowalla, P. M. Ajayan, L. Song, *Chem. Mater.* **2017**, *29*, 4738.
[100] M. Boudart, *Chem. Rev.* **1995**, *95*, 661.
[101] T. Shinagawa, A. T. Garcia-Esparza, K. Takanabe, *Sci. Rep.* **2015**, *5*, 13801.
[102] L. Najafi, S. Bellani, R. Oropesa-Nuñez, M. Prato, B. Martín-García, R. Brescia, F. Bonaccorso, *ACS Nano* **2019**, *13*, 3162.
[103] Y. Liu, J. Wu, K. P. Hackenberg, J. Zhang, Y. M. Wang, Y. Yang, K. Keyshar, J. Gu, T. Ogitsu, R. Vajtai, J. Lou, P. M. Ajayan, B. C. Wood, B. I. Yakobson, *Nat. Energy* **2017**, *2*, 17127.
[104] J. Zhang, J. Wu, X. Zou, K. Hackenberg, W. Zhou, W. Chen, J. Yuan, K. Keyshar, G. Gupta, A. Mohite, P. M. Ajayan, J. Lou, *Mater. Today* **2019**, *25*, 28.
[105] L. Najafi, S. Bellani, R. Oropesa-Nunez, B. Martin-Garcia, M. Prato, F. Bonaccorso, *ACS Appl. Energy Mater.* **2019**, *2*, 5373.
[106] J. Shi, X. Wang, S. Zhang, L. Xiao, Y. Huan, Y. Gong, Z. Zhang, Y. Li, X. Zhou, M. Hong, Q. Fang, Q. Zhang, X. Liu, L. Gu, Z. Liu, Y. Zhang, *Nat. Commun.* **2017**, *8*, DOI 10.1038/s41467-017-01089-z.
[107] H. Li, C. Tsai, A. L. Koh, L. Cai, A. W. Contryman, A. H. Fragapane, J. Zhao, H. S. Han, H. C. Manoharan, F. Abild-Pedersen, J. K. Nørskov, X. Zheng, *Nat. Mater.* **2015**, *15*, 48.
[108] G. Ye, Y. Gong, J. Lin, B. Li, Y. He, S. T. Pantelides, W. Zhou, R. Vajtai, P. M. Ajayan, *Nano Lett.* **2016**, *16*, 1097.
[109] S. Bolar, S. Shit, J. S. Kumar, N. C. Murmu, R. S. Ganesh, H. Inokawa, T. Kuila, *Appl. Catal. B Environ.* **2019**, *254*, 432.
[110] L. Cheng, W. Huang, Q. Gong, C. Liu, Z. Liu, Y. Li, H. Dai, *Angew. Chem. Int. Ed.* **2014**, *53*, 7860.
[111] R. Zhang, X. Wang, S. Yu, T. Wen, X. Zhu, F. Yang, X. Sun, X. Wang, W. Hu, *Adv. Mater.* **2017**, *29*, 1605502.

[112] J. Deng, H. Li, S. Wang, D. Ding, M. Chen, C. Liu, Z. Tian, K. S. Novoselov, C. Ma, D. Deng, X. Bao, *Nat. Commun.* **2017**, *8*, 14430.
[113] Y. Li, H. Wang, L. Xie, Y. Liang, G. Hong, H. Dai, *J. Am. Chem. Soc.* **2011**, *133*, 7296.
[114] J. D. Holladay, J. Hu, D. L. King, Y. Wang, *Catal. Today* **2009**, *139*, 244.
[115] K. Zeng, D. Zhang, *Prog. Energy Combust. Sci.* **2010**, *36*, 307.
[116] F. Safizadeh, E. Ghali, G. Houlachi, *Int. J. Hydrog. Energy* **2015**, *40*, 256.
[117] I. Moussallem, J. Jörissen, U. Kunz, S. Pinnow, T. Turek, *J. Appl. Electrochem.* **2008**, *38*, 1177.
[118] D. Pletcher, X. Li, *Int. J. Hydrog. Energy* **2011**, *36*, 15089.
[119] M. Carmo, D. L. Fritz, J. Mergel, D. Stolten, *Int. J. Hydrog. Energy* **2013**, *38*, 4901.
[120] S. A. Grigoriev, V. I. Porembsky, V. N. Fateev, *Int. J. Hydrog. Energy* **2006**, *31*, 171.
[121] F. Qi, X. Wang, B. Zheng, Y. Chen, B. Yu, J. Zhou, J. He, P. Li, W. Zhang, Y. Li, *Electrochimica Acta* **2017**, *224*, 593.
[122] C. Gammer, C. Mangler, C. Rentenberger, H. P. Karnthaler, *Scr. Mater.* **2010**, *63*, 312.
[123] L. Wang, Z. Sofer, J. Luxa, D. Sedmidubský, A. Ambrosi, M. Pumera, *Electrochem. Commun.* **2016**, *63*, 39.

**The table of contents entry:**

**Colloidal synthesis of rhenium disulfide nanosheets enables a simple and cost-effective exploitation of its peculiar layer-independent properties for gas-sensing and electrochemical $H_2$ production**. The surface functionalization of the nanosheets leads to sensitive and fast response gas sensors, while their assembly with carbon nanotubes enhances its electrocatalytic activity, making both device performance competitive with CVD rhenium disulfide.

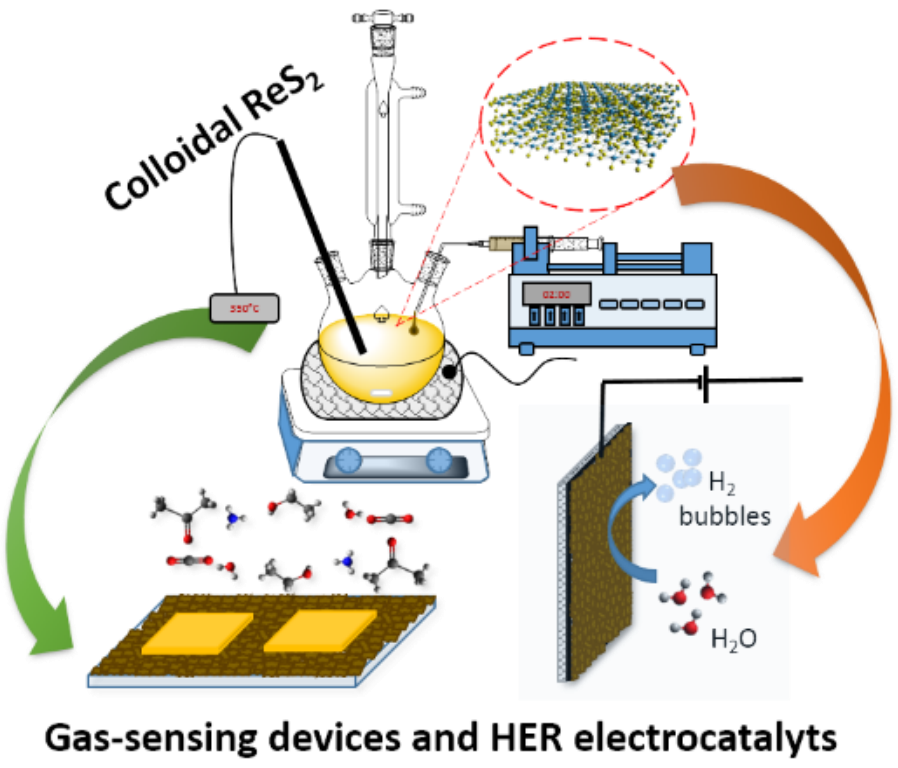

# Supporting Information

## Additional c-$ReS_2$ material characterization

In contrast to the high-magnification transmission electron microscopy (TEM) images displayed in Figure 1a-b of the main text, which allow us to discern individual colloidal $ReS_2$ (c-$ReS_2$) nanosheets, low-magnification TEM images in **Figure S1a**, only show assemblies of nanosheets. We also found these assemblies formed by the c-$ReS_2$ nanosheets as well as small aggregates (lateral size <100 nm), with thickness from 6 to 3 nm by atomic force microscopy (AFM) analysis, reported in **Figure S1b-c.** The height-profile statistical analysis of the small assemblies of c-$ReS_2$ nanosheets (performed by excluding big aggregates) shows a mean value of 12.2 nm and a median of 7.8 nm. The data follow a lognormal distribution peaked at 4.3 nm with 23.8 % of nanosheets. **Figure S1d** shows the estimation of the band gap ($E_g$) from $(\alpha h\nu)^n$ *vs.* $h\nu$ (Tauc plot) analysis using the Tauc relation $Ah\nu = Y(h\nu\text{-}E_g)^n$. With $n=2$ ($ReS_2$ is a direct-gap semiconductor), we find an $E_g$ value of about 1.41 eV for the colloidal nanosheets.

We carried out elemental analysis of the c-$ReS_2$ by means of scanning electron microscopy (SEM) and energy dispersive X-Ray spectroscopy (EDS) in a Helios Nanolab 600 (FEI Company) microscope combined with an X-Max detector and INCA® system (Oxford Instruments). For the EDS spectra acquisition and analysis measurement conditions were set at 15kV and 0.8nA. Results shown a stoichiometry ratio Re:S of 1:1.4 from the analysis of the Re and S peaks (**Figure S1e**). As comparison, we also performed the SEM/EDS analysis in the liquid phase exfoliated (LPE) and bulk counterparts obtaining a stoichiometry ratio Re:S of 1:1.5 and 1:1.6, respectively. The values estimated in all cases are smaller than the expected 1:2 ratio for $ReS_2$, and in agreement with the X-Ray photoelectron spectroscopy (XPS) data presented in the main text (Figure 2a-b).

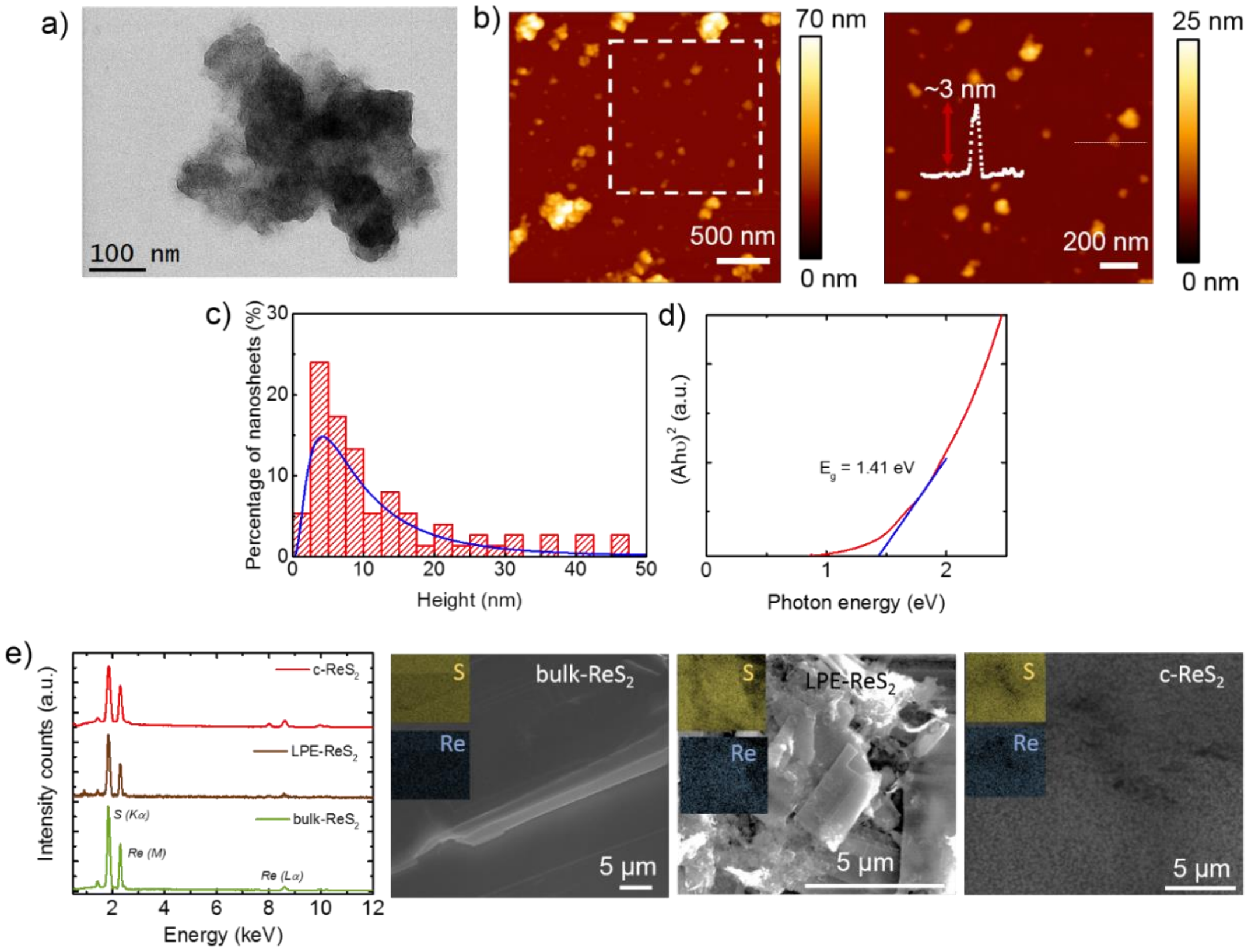

**Figure S1.** (a) Representative TEM image of the c-$ReS_2$ nanosheets. (b-c) Representative AFM images and heigh-profile statistics carried out on 75 assemblies of c-$ReS_2$ nanosheets (excluding big aggregates). The lognormal fit

(blue line) is also shown. The white-dashed square in the AFM image shown on the left corresponds to the more resolved AFM scan shown on the right. (d) Tauc plot of c-$ReS_2$ with the corresponding linear fit for the estimation of the $E_g$. (e) SEM/EDS spectra and mapping analysis for the Re (M = 1.84 keV and Lα = 8.65 keV) and S (Kα = 2.31 keV) lines of c-$ReS_2$, LPE-$ReS_2$ and bulk $ReS_2$ samples drop-cast on copper substrates (Cu, Lα = 0.93 keV and Kα = 8.04 keV).

## Liquid phase exfoliation of $ReS_2$ and characterization

$ReS_2$ flakes (**Figure S2a-b**) were produced from bulk $ReS_2$ crystals (HQGraphene®) by liquid phase exfoliation in isopropanol (IsOH) followed by sedimentation-based separation.[1–3] Briefly 20 mg of bulk $ReS_2$ was added to 5 mL of IsOH and then, ultrasonicated (Branson® 5800 cleaner, Branson ultrasonics) for 6h. The resulting dispersion was centrifuged for 20 min at 935*g* (Sigma 3-16P centrifuge, rotor 19776), in order to separate the exfoliated $ReS_2$ flakes as supernatant from the un-exfoliated and thick $ReS_2$ crystals remaining on the bottom of the vial. Only 80% of the supernatant volume is collected by pipetting, avoiding possible contamination from the precipitated powder. **Figure S2c** shows the estimation of the $E_g$ from $(\alpha h\nu)^n$ *vs.* $h\nu$ (Tauc plot) analysis,[4] resulting in an $E_g$ value of about 1.43 eV.

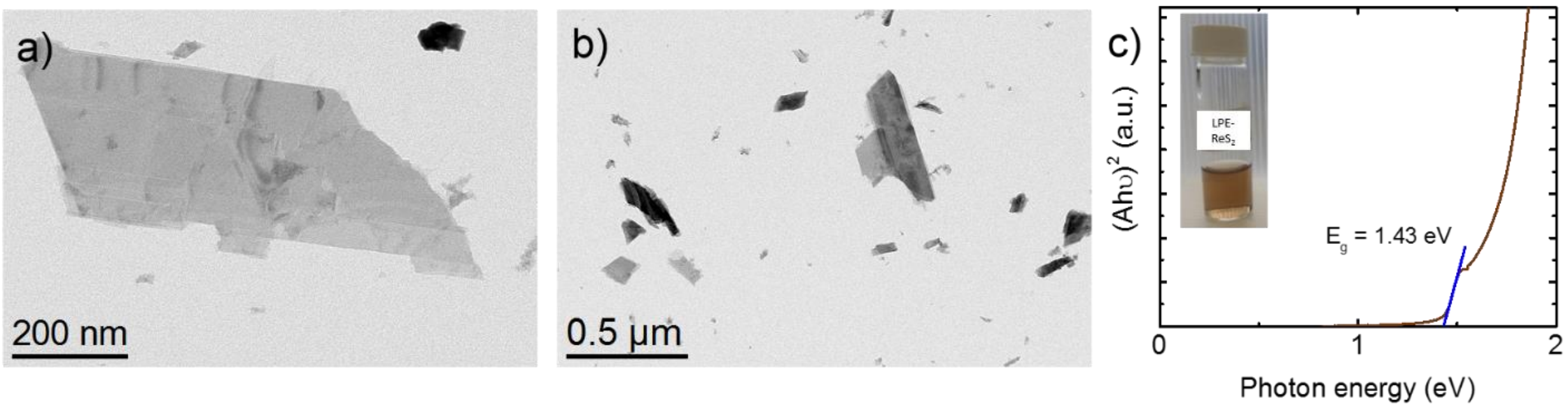

**Figure S2.** (a,b) Representative TEM images of the LPE-$ReS_2$ flakes. (c) Tauc plot of the LPE-$ReS_2$ sample with the corresponding fitting for the estimation of the $E_g$. The inset shows a photograph of the LPE-$ReS_2$ dispersion in IsOH.

## Supplementary SAED information

As mentioned in the main text, we evaluate the crystal structure of the samples from the TEM/selected-area electron diffraction (SAED) analysis. From the TEM images and the corresponding selected-area diffraction (SAED) patterns shown in **Figure S3a-d** for c-$ReS_2$ and LPE-$ReS_2$ samples, we obtained the background-subtracted and azimuthally integrated SAED patterns displayed in **Figure S3e**.

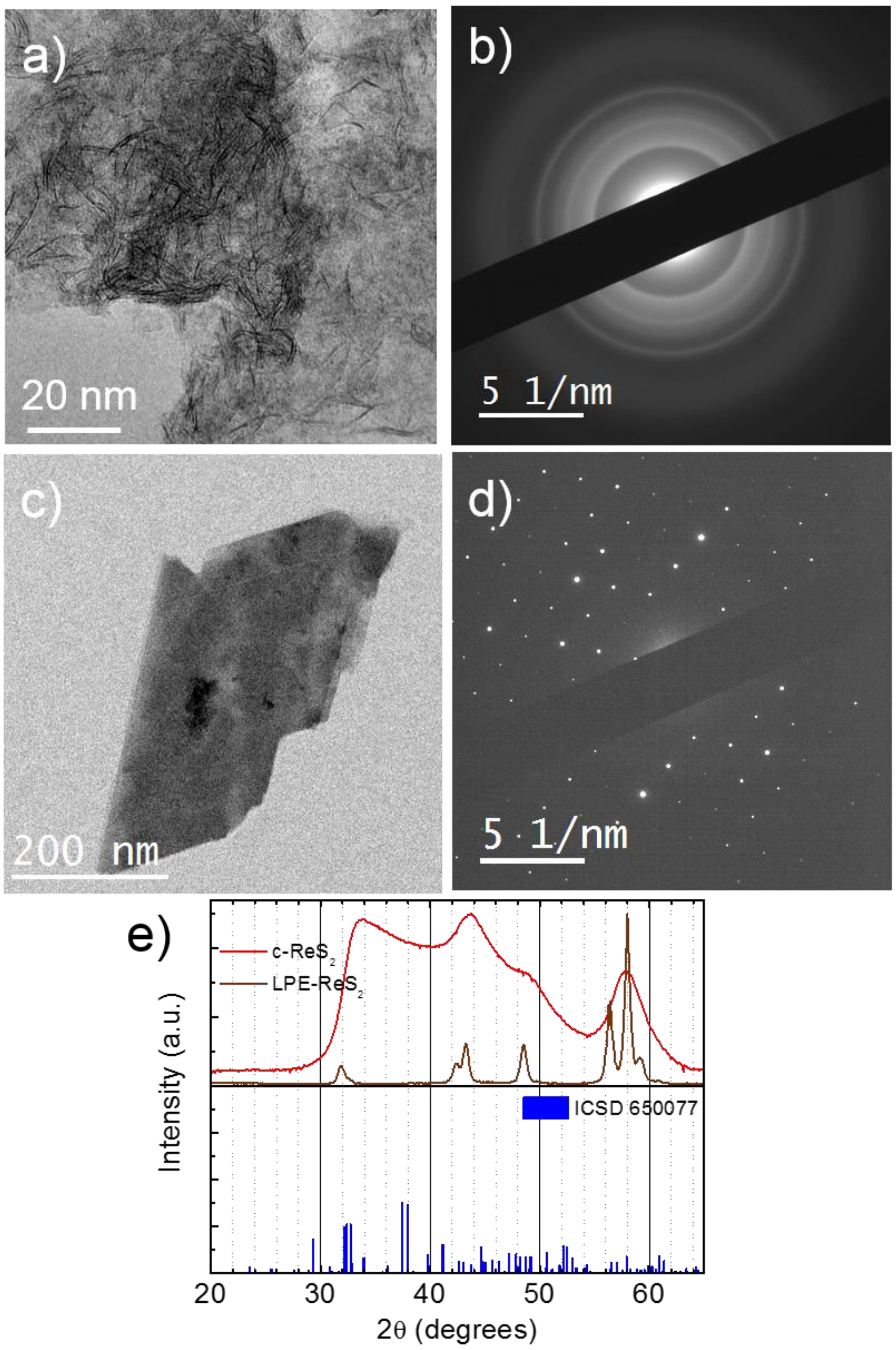

**Figure S3.** (a, c) BF-TEM images and (b, c) corresponding SAED patterns of (a, b) an aggregate of c-$ReS_2$ nanosheets and (c, d) a flake in the LPE-$ReS_2$ sample. (e) The background-subtracted and azimuthally integrated SAED patterns, plotted as a function of the Bragg angle calculated for Cu Kα wavelength, are compared to the powder XRD pattern for triclinic $ReS_2$ (ICSD structure n. 650077).

## Discussion of the $ReS_2$ Raman spectra

As mentioned in the main text regarding **Figure 1d-e** , the analysis of the Raman spectra has been carried out following the detailed reports on $ReS_2$ performed by Balicas' and Terrones' groups.[5,6] Therefore, the in-plane mode near 150 $cm^{-1}$ is labeled as $A_g^1$, the out-of-plane mode around 437 $cm^{-1}$, where just the S atoms vibrate, is assigned as $A_g^2$, and a quasi-out-of-plane mode that is denoted as $A_g^3$ can be identified around 418 $cm^{-1}$, in which the S vibrations occur at a small angle with respect to the basal plane. The remaining modes, in which the atoms vibrate at different angles due to the low symmetry of $ReS_2$, are denoted as $A_g^4$ to $A_g^{18}$ with increasing frequency. With a 532 nm excitation wavelength, Figure 1e (main text), we observe the main Raman modes.[6] In the c-$ReS_2$ , the Re in-plane vibrations ($A_g^1$ 151 $cm^{-1}$, $A_g^7$ 212 $cm^{-1}$, $A_g^8$ 237 $cm^{-1}$) and S vibrations ($A_g^{15}$ 349 $cm^{-1}$, $A_g^{18}$ 408 $cm^{-1}$, $A_g^2$ 436 $cm^{-1}$) are clearly visible. The LPE sample shows a similar behavior, in which Re in-plane vibrations ($A_g^1$ 152 $cm^{-1}$; $A_g^6$ 162 $cm^{-1}$; $A_g^7$ 213 $cm^{-1}$; $A_g^8$ 235 $cm^{-1}$) and S vibrations ($A_g^{15}$ 346 $cm^{-1}$, $A_g^{18}$ at 407 $cm^{-1}$, $A_g^3$ 418 $cm^{-1}$, $A_g^2$ 437 $cm^{-1}$) are clearly detected, but shows also two additional modes ($A_g^6$ 162 cm-

[1]; and $A_g^3$ 418 $cm^{-1}$) compared to the c-$ReS_2$ that are probably due to the higher crystallinity of the LPE sample.

**Characterization of colloidal $ReSe_2$ nanosheets**

As introduced in the main text, colloidal $ReSe_2$ nanosheets (c-$ReSe_2$) can be obtained from the same synthetic protocol than for the c-$ReS_2$ by changing S powder for Se powder (86 mg, 99.99%, Strem Chemicals®) using the same $ReCl_5$ (67 mg) as Re precursor, and setting the temperature to 350°C. As happened for $ReS_2$, this temperature is lower than the 625°C established for the respective CVD approaches.[7] For comparative purposes, we prepared the corresponding LPE counterpart in IsOH following the same protocol described for $ReS_2$ exfoliating the $ReSe_2$ crystal supplied from HQGraphene®.

The $ReSe_2$ nanosheets obtained are assembled in form of flowers as can be observed in the corresponding low-magnification TEM images (**Figure S4a**). From high-magnification TEM images, a lateral size of ca. 4 ± 1 nm and thickness of 0.4 ± 0.1 nm is estimated for the tilted c-$ReSe_2$ nanosheets. As happened for c-$ReS_2$ nanosheets, this value is lower than the one reported for CVD-growth hexagonal $ReSe_2$ monolayer (0.7 nm)[7] or CVD-growth distorted 1T $ReSe_2$ monolayer (0.9 nm),[8] probably due to the heavy strain building up in these highly anisotropic 2D colloidal structures.[9] Moreover, the SAED patterns from c-$ReSe_2$ nanosheets and LPE-$ReSe_2$ flakes (**Figure S4c**) indicate comparable 2θ values for Bragg peak positions, corresponding to the hexagonal structure. As for the c-$ReS_2$, much broader peaks characterize the c-$ReSe_2$ nanosheets, due to the much smaller size of single-crystal domains (few nm in the colloidal sample vs. hundreds of nm for LPE flakes). From the SEM/EDS analysis, **Figure S4d**, we estimated a Re:Se ratio of 1:1.5 (c-$ReSe_2$), 1:1.5 (LPE-$ReSe_2$) and 1:1.6 (bulk-$ReSe_2$) from the Re and Se peaks. This ratio is slightly smaller than would be expected from the 1:2 $ReSe_2$ stoichiometry, and in accordance with the elemental analysis data obtained also for the $ReS_2$ samples. We also evaluated the extinction coefficient and Raman spectra that are plotted in **Figure S4e** and **f**. As expected for $ReSe_2$, the material shows a strong and broad extinction from 300 nm to almost 1000 nm. However, in contrast to its LPE equivalent, there is no clear excitonic peak at ~920 nm that corresponds to the $E_g$ of the $ReSe_2$.[10] The Raman spectra in **Figure S4e** shows the presence of the characteristics Raman modes of $ReSe_2$ in the 100 to 300 $cm^{-1}$ range. Since as reported by Wolverson and collaborators,[11] the relative intensities of the bands are dependent on the orientation of the flakes and do not scale proportionately with thickness, we just carried out a qualitative interpretation of the Raman spectrum of c-$ReSe_2$ in comparison with its LPE and bulk counterparts. As can be seen in Figure S4e, the Raman peaks from the colloidal sample are broader than the corresponding ones in the LPE sample, indicating, as for the c-$ReS_2$ case, a lower degree of crystallinity compared to the LPE one.[12] The low crystallinity of the c-$ReSe_2$ nanosheets might also be responsible for the absence of the excitonic peak in the spectrum in Figure S4d, and attributed to the lower synthesis temperature in the colloidal approach (350°C) in comparison with CVD approach (625°C [7]).

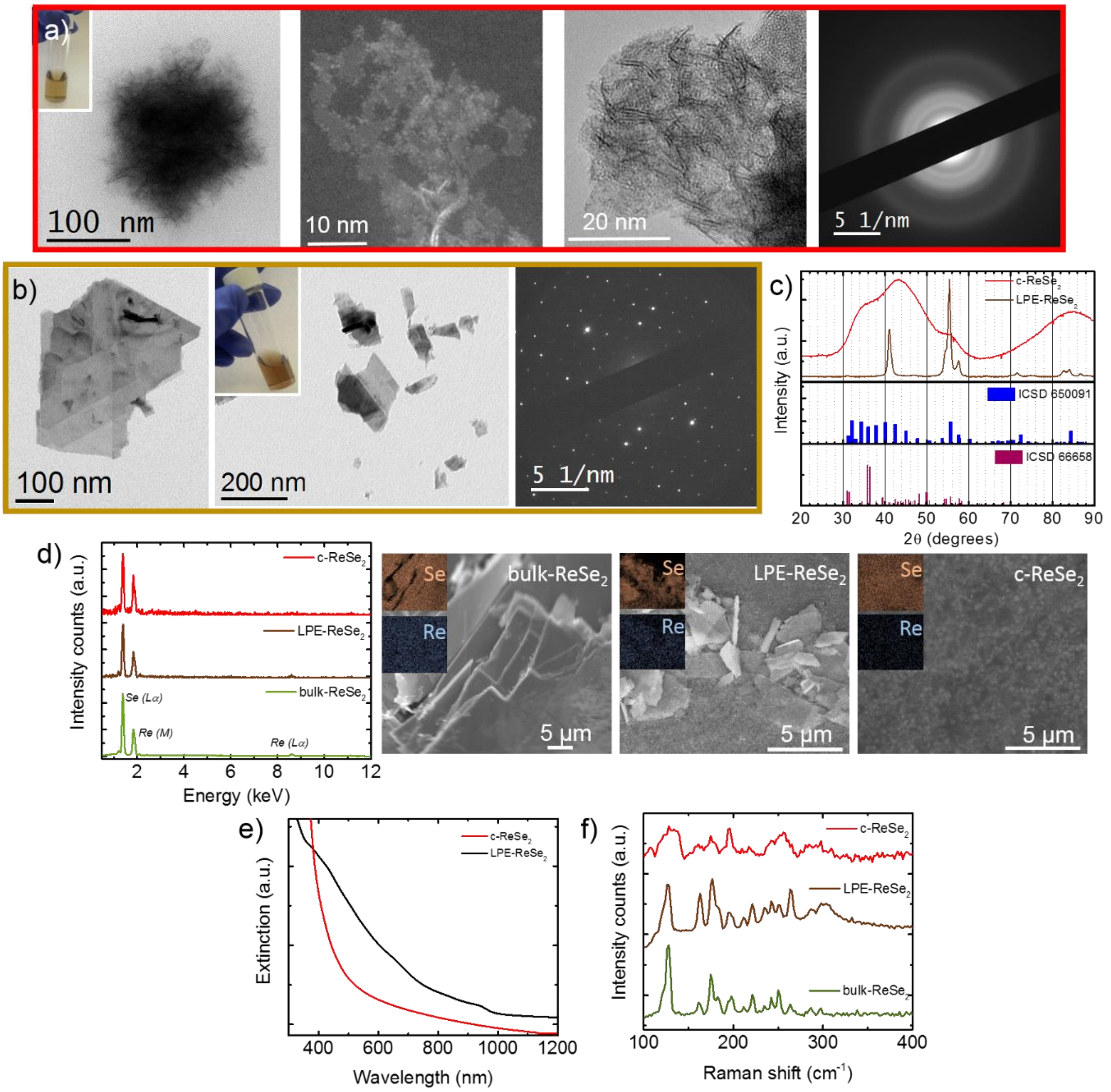

**Figure S4.** (a,b) Representative TEM and HAADF-STEM images of the c-$ReSe_2$ sheets (red square) and LPE-$ReSe_2$ flakes (brown square), respectively, together with the corresponding selected-area diffraction (SAED) patterns at the end. (c) The background-subtracted and azimuthally integrated SAED patterns, plotted as a function of the Bragg angle calculated for Cu Kα wavelength, are compared to the powder XRD pattern for triclinic $ReSe_2$ (ICSD structure n. 66658) and hexagonal $ReSe_2$ (ICSD structure n. 650091). The insets show the corresponding photographs of the c-$ReSe_2$ dispersion in toluene and LPE-$ReSe_2$ dispersion in isopropanol. (d) SEM/EDS spectra and mapping analysis for the Re (M = 1.84 keV and Lα = 8.65 keV) and Se (Lα = 1.38 keV) lines of c-$ReSe_2$, LPE-$ReSe_2$ and bulk $ReSe_2$ samples drop-cast on Au-coated glass (Au, M = 2.12 keV and Lα = 9.71 keV). (e) Extinction spectra of c-$ReSe_2$ and LPE-$ReSe_2$ dispersed in toluene and isopropanol, respectively. An offset was applied to the curves for clarity. (f) Representative Raman spectra collected at 532 nm excitation wavelength on the c-$ReSe_2$, LPE-$ReSe_2$ and bulk $ReSe_2$ samples drop-cast on silicon wafers.

## Gas sensors: I-V curves, SEM-EDX supplementary characterization, device performance for functionalized c-$ReS_2$

For comparison, the I-V characteristics under inert $N_2$ of the different gas sensing devices prepared were collected as shown in **Figure S5a-b**.

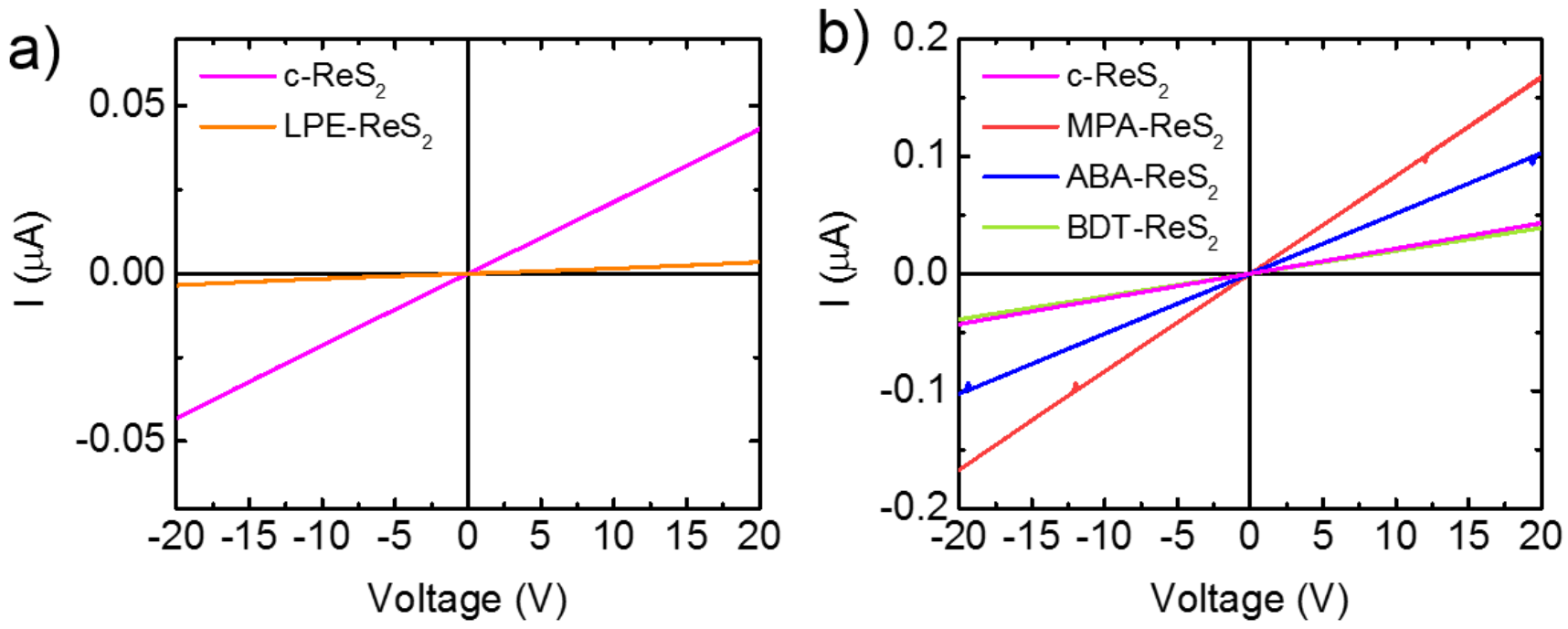

**Figure S5.** (a) I-V characteristics recorded under inert $N_2$ of the different devices prepared from c-$ReS_2$ (magenta) and LPE-$ReS_2$ (orange). (b) I-V characteristics acquired under inert $N_2$ of the different devices for different $ReS_2$ nanosheet functionalization: MPA (red); ABA (blue) and BDT (green).

In order to obtain films without holes or cracks for the fabrication of the gas-sensing devices, we tested different spin-coating and drop-casting procedures. Scanning electron microscopy (SEM) images in **Figure S6a** show the formation of holes in the film prepared with MPA-exchanged c-$ReS_2$ by using spin coating, which can affect the sensor performance. Therefore, we prepared the films by drop-casting. With this approach we obtained a more homogeneous films independent of the functionalization molecule as shown in the SEM images in **Figure S6b-c** for devices built with ABA- and BDT- exchanged c-$ReS_2$, respectively.

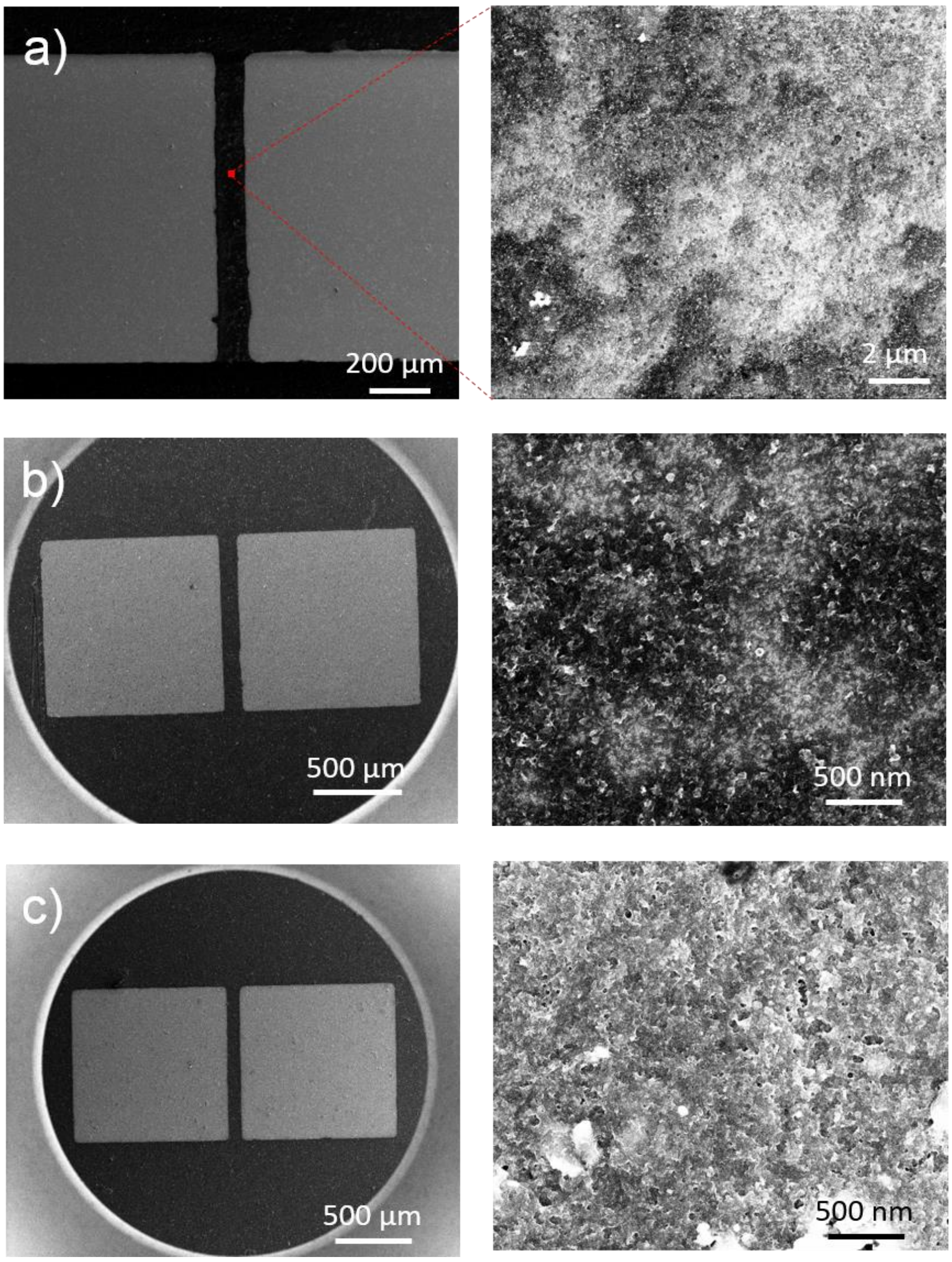

**Figure S6.** (a) Representative SEM images of a device fabricated with MPA-exchanged c-$ReS_2$, in which the film was prepared by spin coating method highlighting the formation of holes in the film. (b, c) Representative SEM images of gas sensors built with ABA- and BDT-exchanged c-$ReS_2$, in which the film was prepared by drop-casting method showing a homogeneous film (thickness of about 400 nm), respectively.

**Table S1** summarizes the performance of the different gas sensors in terms of conductance (G)[13] and resistance (R) response. **Table S2** reports the response and recovery times determined as the time needed to reach 90% of response in the presence of gases, and to return to 10% of the response, respectively.[13] For the recovery time measurement, a continuous $N_2$ flow was applied to purge the gas chamber.

**Table S1.** Averaged response calculated as $(G_{gas} - G_{inert})/G_{inert}$ and $(R_{gas} - R_{inert})/R_{inert}$, for different devices and gases averaged over several repetitions of on/off cycles.

| Device | Conductance response | | | | | |
|---|---|---|---|---|---|---|
| | $NH_3$ | $H_20$ | $CO_2$ | EtOH | acetone | air |
| **ann.-c-ReS$_2$** | 2.4±0.2 | 1.2±0.1 | | -0.15±0.01 | -0.12±0.02 | -0.061±0.003 |
| **LPE 1st pulse** | 4.6*$10^4$ | 1.1*$10^3$ | | | | |
| **LPE following pulses** | 9.9 | 8.8 | | | | |
| **MPA-ReS$_2$** | 31±8 | 4.5±2.2 | 0.8±0.4 | -0.18±0.01 | -0.12±0.01 | -0.034±0.001 |
| **BDT-ReS$_2$** | 24±3 | 2.4±0.5 | | | | |
| **ABA-ReS$_2$** | 16±2 | 2.6±0.3 | | | | |

| Device | Resistance response | | | | | |
|---|---|---|---|---|---|---|
| | $NH_3$ | $H_20$ | $CO_2$ | EtOH | acetone | air |
| **ann.-c-ReS$_2$** | -0.70±0.01 | -0.54±0.03 | | 0.18±0.01 | 0.14±0.03 | 0.064±0.003 |
| **LPE 1st pulse** | -0.99998 | -0.9991 | | | | |
| **LPE following pulses** | -0.90 | -0.891 | | | | |
| **MPA-ReS$_2$** | -0.97±0.01 | -0.78±0.08 | -0.42±0.12 | 0.21±0.01 | 0.14±0.01 | 0.003±0.001 |
| **BDT-ReS$_2$** | -0.96±0.04 | -0.70±0.04 | | | | |
| **ABA-ReS$_2$** | -0.93±0.01 | -0.71±0.02 | | | | |

**Table S2.** Rise ($\tau_R$) and fall ($\tau_F$) times of the conductance or resistance response for the different devices and gases.

| Device | $\tau_R$ (s) | | | | | | $\tau_F$ (s) | | | | | |
|---|---|---|---|---|---|---|---|---|---|---|---|---|
| | $NH_3$ | $H_20$ | $CO_2$ | EtOH | acetone | air | $NH_3$ | $H_20$ | $CO_2$ | EtOH | acetone | air |
| **ann.-c-ReS$_2$** | 9.3 | 8.7 | | 5.2 | 1.0 | 2.3 | 1.3 | 0.3 | | 25 | 56 | 12 |
| **LPE** | 5.2 | 5.2 | | | | | 30 | 31 | | | | |
| **MPA-ReS$_2$** | 3.3 | 4.1 | 6.5 | 4.7 | 2.3 | 2.8 | 1.0 | 0.5 | 1.0 | 9.0 | 6.9 | 5.6 |
| **BDT-ReS$_2$** | 6.5 | 5.7 | | | | | 0.8 | 1.0 | | | | |
| **ABA-ReS$_2$** | 8.0 | 8.0 | | | | | 1.3 | 1.8 | | | | |

In order to estimate the minimum amount of gas ($NH_3$ or $H_2O$) detectable according with the design of the technique, we have calculated the minimum concentration (amount of gas) that we can detect taking into account the noise level of our device in terms of current read ($I_{noise}$). Therefore, we consider that, for the current corresponding to the minimum amount of gas $I_{gas,min}$, we have

$$I_{gas,min} - I_{inert} \geq I_{noise}$$

$I_{noise}$, was estimated by measuring the current under inert atmosphere for 10 s ($I_{inert}$) and calculating the standard deviation. Then, we assume a linear trend of the response with the concentration of gas (Q); that is,

$$I_{gas} - I_{inert} = \alpha Q$$

and the sensitivity α can be calculated from the measurement reported in Figures 2d-f and 3c-d and in the SI, Table S1 as:

$$\alpha = \frac{I_{gas} - I_{inert}}{Q_0}$$

Where $Q_0$ is the reference gas concentration, which for $H_2O$ is 95% relative humidity (measured with a humidity sensor at the inlet, and equivalent to 23080 ppm on volume[14]), while for we use a $NH_3$ aqueous solution at 2% in volume (thus, ca. 500 ppm).

With these assumptions, the minimum gas quantity $Q_{min}$ can be determined as:

$$Q_{min} = \frac{I_{noise}}{\alpha}$$

In the Tables S3 and S4, it is reported the $Q_{min}$ values for $NH_3$ and $H_2O$ detectable for the different devices according with the design of the technique.

**Table S3.** Minimum gas quantity ($Q_{min}$) for $NH_3$ detectable for the different devices according with the technique of measurement.

| Device – $NH_3$ detection | α(A/ppm) | $I_{noise}$ (A) | $Q_{min}$ (ppm) |
|---|---|---|---|
| ann.-c-$ReS_2$ | $5.2\cdot\cdot 10^{-10}$ | $4.1\cdot 10^{-11}$ | 0.08 |
| LPE 1st pulse | $3.4\cdot 10^{-7}$ | $8.7\cdot 10^{-10}$ | 0.003 |
| LPE following pulses | $2.9\cdot 10^{-7}$ | $6.1\cdot 10^{-7}$ | 2.1 |
| MPA-$ReS_2$ | $2.7\cdot 10^{-8}$ | $3.0\cdot 10^{-1\text{ß}}$ | 0.01 |
| BDT-$ReS_2$ | $1.8\cdot 10^{-9}$ | $1.3\cdot 10^{-11}$ | 0.007 |
| ABA-$ReS_2$ | $3.5\cdot 10^{-9}$ | $2.2\cdot 10^{-10}$ | 0.06 |

**Table S4.** Minimum gas quantity ($Q_{min}$) for $H_2O$ detectable for the different devices according with the technique of measurement.

| Device – $H_2O$ detection | α(A/ppm) | $I_{noise}$ (A) | $Q_{min}$ (ppm) |
|---|---|---|---|
| ann.-c-$ReS_2$ | $5.8\cdot 10^{-12}$ | $4.1\cdot 10^{-11}$ | 7.1 |
| LPE 1st pulse | $1.8\cdot 10^{-10}$ | $8.7\cdot 10^{-10}$ | 5.0 |
| LPE following pulses | $5.7\cdot 10^{-10}$ | $6.1\cdot 10^{-7}$ | 110 |
| MPA-$ReS_2$ | $8.6\cdot 10^{-11}$ | $3.0\cdot 10^{-1\text{ß}}$ | 3.5 |
| BDT-$ReS_2$ | $3.9\cdot 10^{-11}$ | $1.3\cdot 10^{-11}$ | 3.3 |
| ABA-$ReS_2$ | $1.2\cdot 10^{-11}$ | $2.2\cdot 10^{-10}$ | 18 |

For the stability-retention tests, the measurements comprised 3 steps: (i) inert ($N_2$) gas flow; (ii) target gas flow of about 10s; and (iii) flow paused during 10-15s, while the device remains exposed under an environment of the target gas (indicated as “env”). As can be seen in **Figure S7** for the device fabricated with MPA-exchanged c-$ReS_2$ under different gases: $NH_3$, $H_2O$, EtOH and acetone, the conductance increases under the presence of $NH_3$ and $H_2O$ (or decreases for EtOH or acetone) with the gas inlet, keeping the modification during the first seconds of the flow paused entering in a slight fluctuation due to the absorption/desorption equilibrium. As soon as the $N_2$ flow is again introduced, the desorption of the gas molecules occurs, reflected in the decrease for $NH_3$ and $H_2O$ (or increase for EtOH or acetone) of the conductance up to get back the initial state.

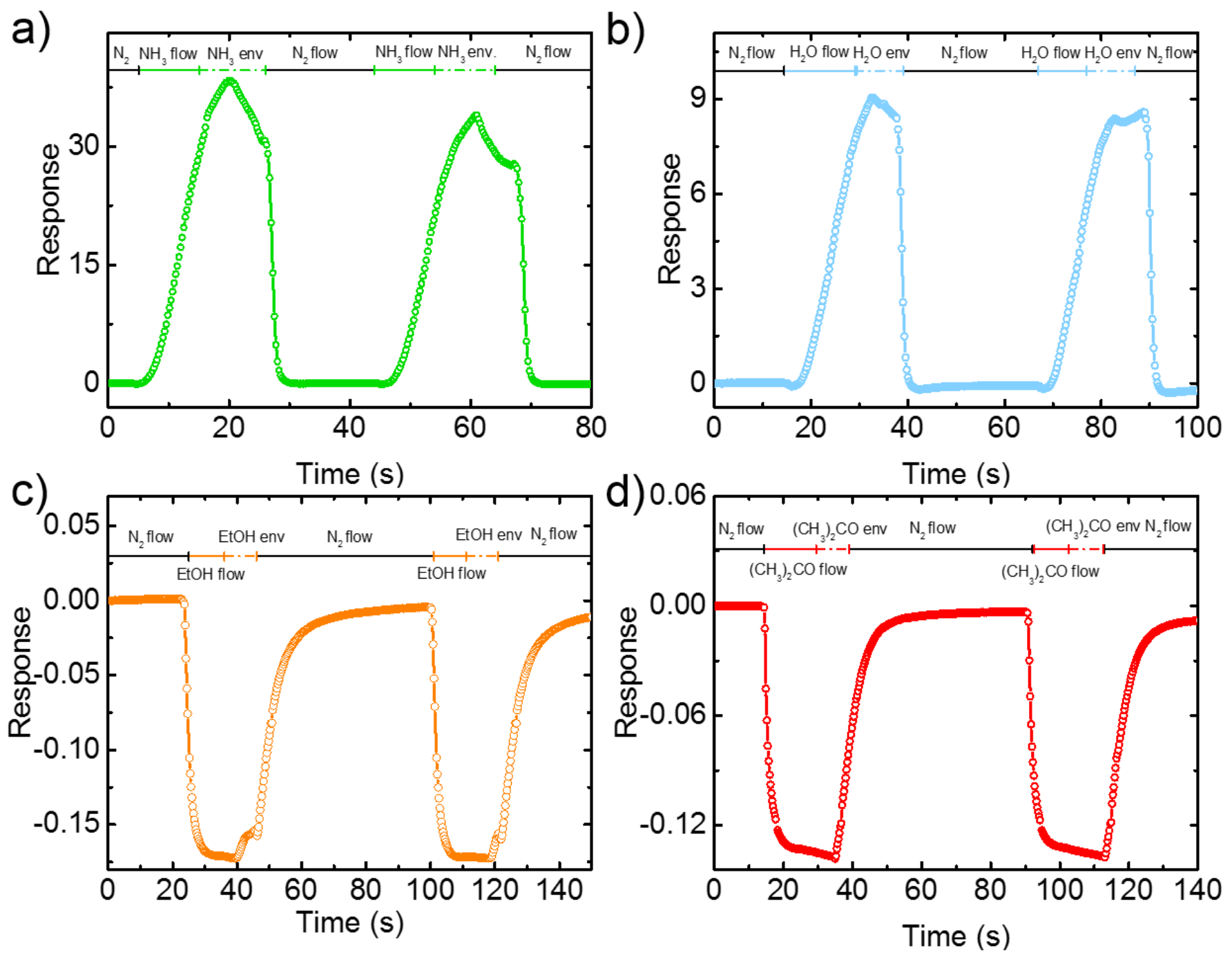

**Figure S7.** Representative stability-retention tests on devices fabricated with MPA-exchanged c-$ReS_2$ under 10s steps for different target gases, (a) $NH_3$; (b) $H_2O$; (c) ethanol and (d) acetone, combining 3 steps: (i) gas inlet; (ii) gas flow paused (noted as 'env.') and (iii) $N_2$ cleaning. The gas-induced response was determined from the film conductance variation as detailed in the main text $[(G_{gas} - G_{inert})/G_{inert}]$.

To demonstrate the sensitivity of the devices built from ABA- and BDT-exchanged c-$ReS_2$, **Figure S8** includes the gas-response determined from the conductance under the exposure of $NH_3$ and $H_2O$. As pointed out in the main text and **Table S1**, the higher response is achieved under the exposure to $NH_3$ gas, in agreement with theoretical calculations on sensors fabricated with $ReS_2$ reported in literature[15].

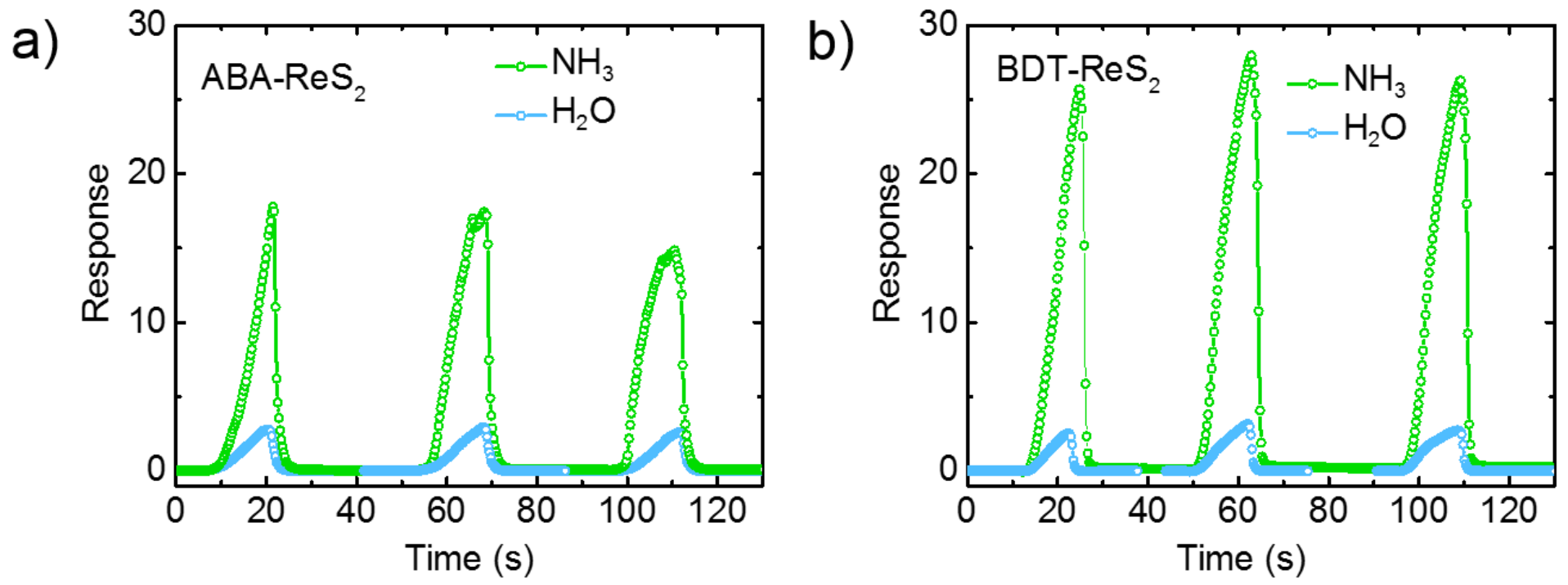

**Figure S8.** Representative response of the devices made with ABA- (a) and BDT- (b) exchanged c-$ReS_2$ under the presence of $NH_3$ and moisture ($H_2O$). The gas-induced response was determined from the film conductance variation as detailed in the main text $[(G_{gas} - G_{inert})/G_{inert}]$.

## Complementary electrodes characterization and electrochemical tests

As indicated in the main text, the SWCNTs/c-$ReS_2$ assemblies for the preparation of the electrodes show a heterogeneous size dispersion of the $ReS_2$ crystals from 0.5 to 200 µm. The big crystals detach from the membrane after the immersion of the electrodes in the electrolyte, not contributing to the HER-activity. However, the $ReS_2$ crystals (< 1 µm) are able to penetrate the SWCNT network as shown in **Figure S9** and participate in the HER.

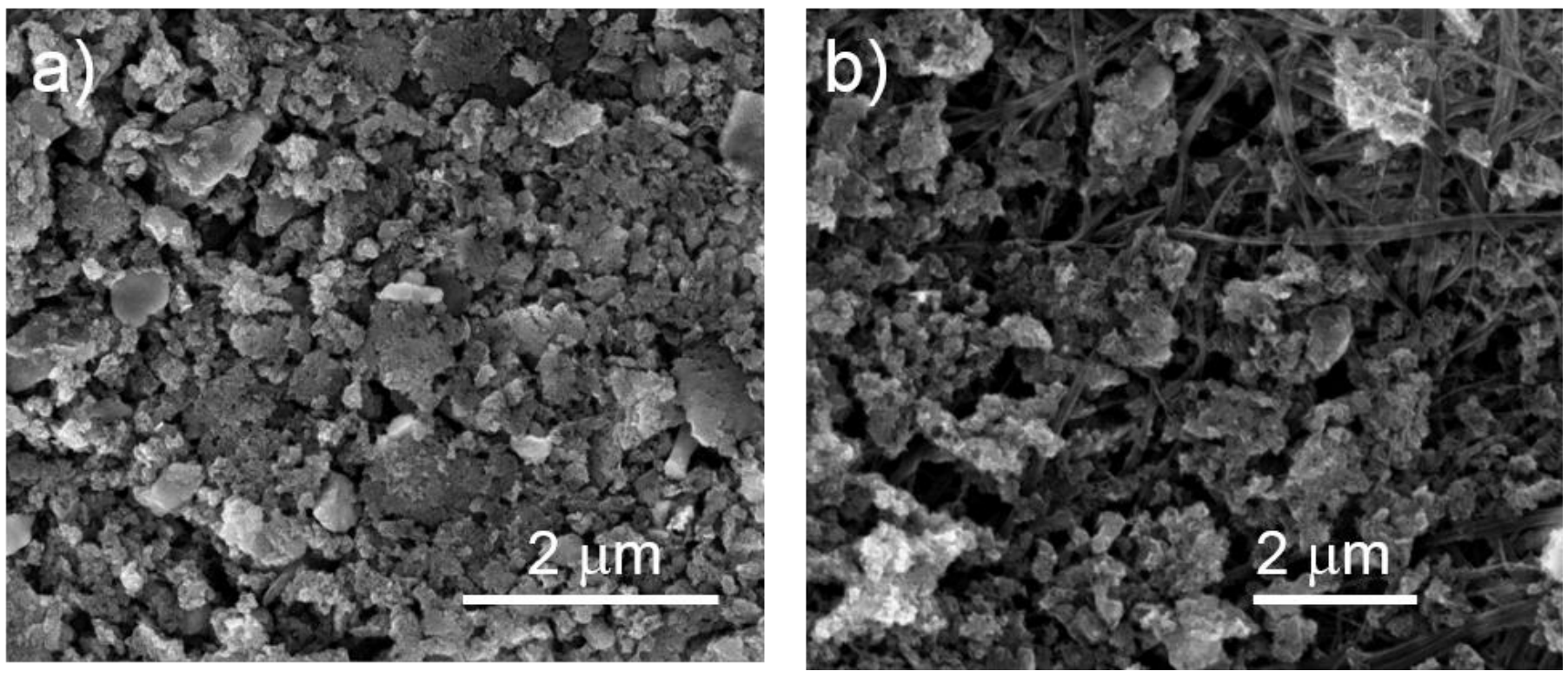

**Figure S9.** Top-view (a) and cross sectional (b) SEM images of SWCNTs/ c-$ReS_2$, revealing a partial penetration of $ReS_2$ crystals into the SWCNT network.

**Figure S10a** shows a top-view photograph of a representative SWCNTs/LPE-$ReS_2$ (electrode area of ~3.5 $cm^2$). **Figures S10b-c** display top-view SEM images of the SWCNTs/LPE-$ReS_2$. The LPE-$ReS_2$ flakes form a film atop the SWCNTs network. In particular, the highest magnification SEM image (**Figure S10c**) shows the presence of $ReS_2$ flakes with lateral dimensions up to ~0.5 µm. **Figure S10d** reports the polarization curves measured for SWCNTs/LPE-$ReS_2$ in both acidic and alkaline media. The data indicate that the electrode display a significant HER-activity in both media, showing $\eta_{10}$ = 0.192 V in 0.5 M $H_2SO_4$ and $\eta_{10}$ = 0.238 V in 1 M KOH.

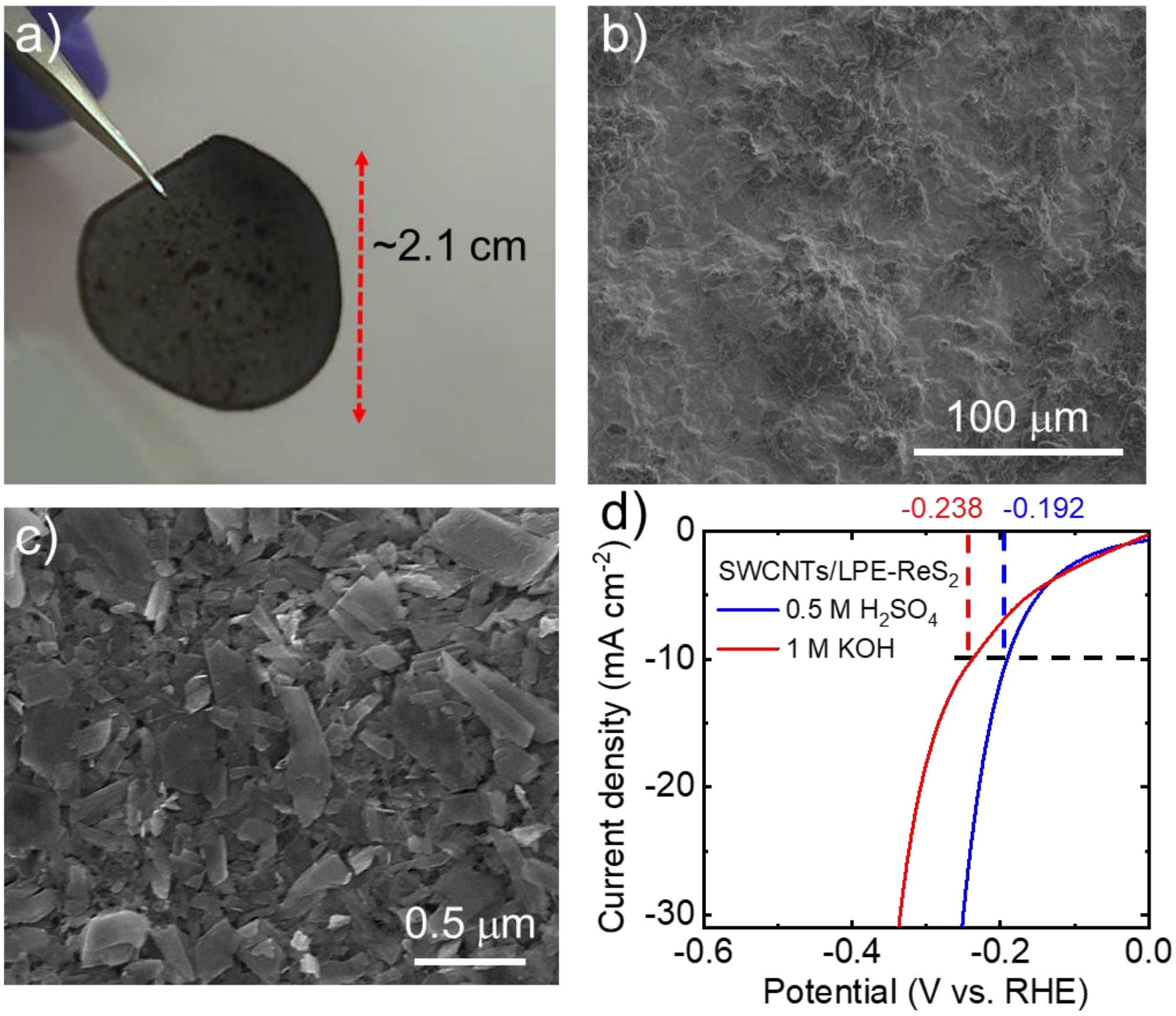

**Figure S10.** (a) Top-view photograph of SWCNTs/LPE-$ReS_2$ (electrode area: ~3.5 $cm^2$). (b-c) Top-view SEM images of SWCNTs/LPE-$ReS_2$. (d) iR-corrected polarization curves of the SWCNTs/LPE-$ReS_2$ in acidic (0.5 M $H_2SO_4$) and alkaline (1 M KOH) media. The $\eta_{10}$ values are also indicated.

**Figure S11** reports the Nyquist plots of SWCNTs/c-$ReS_2$ and SWCNTs/LPE-$ReS_2$ obtained by electrochemical impedance spectroscopy (EIS) measurements at open-circuit potential in 0.5 M $H_2SO_4$ (panel a) and 1 M KOH (panel b). These data do not evidence any significant difference between the electrodes. In addition, the Re[Z] values measured at 100 kHz resemble those one measured by single frequency EIS (in 0.5 M $H_2SO_4$: ~1.4 Ω for SWCNTs/c-$ReS_2$ and ~0.95 Ω for SWCNTs/LPE-$ReS_2$; in 1 M KOH: ~2.2 Ω for SWCNTs/c-$ReS_2$ and ~2.0 Ω for SWCNTs/LPE-$ReS_2$), which was performed before polarization curve measurements.

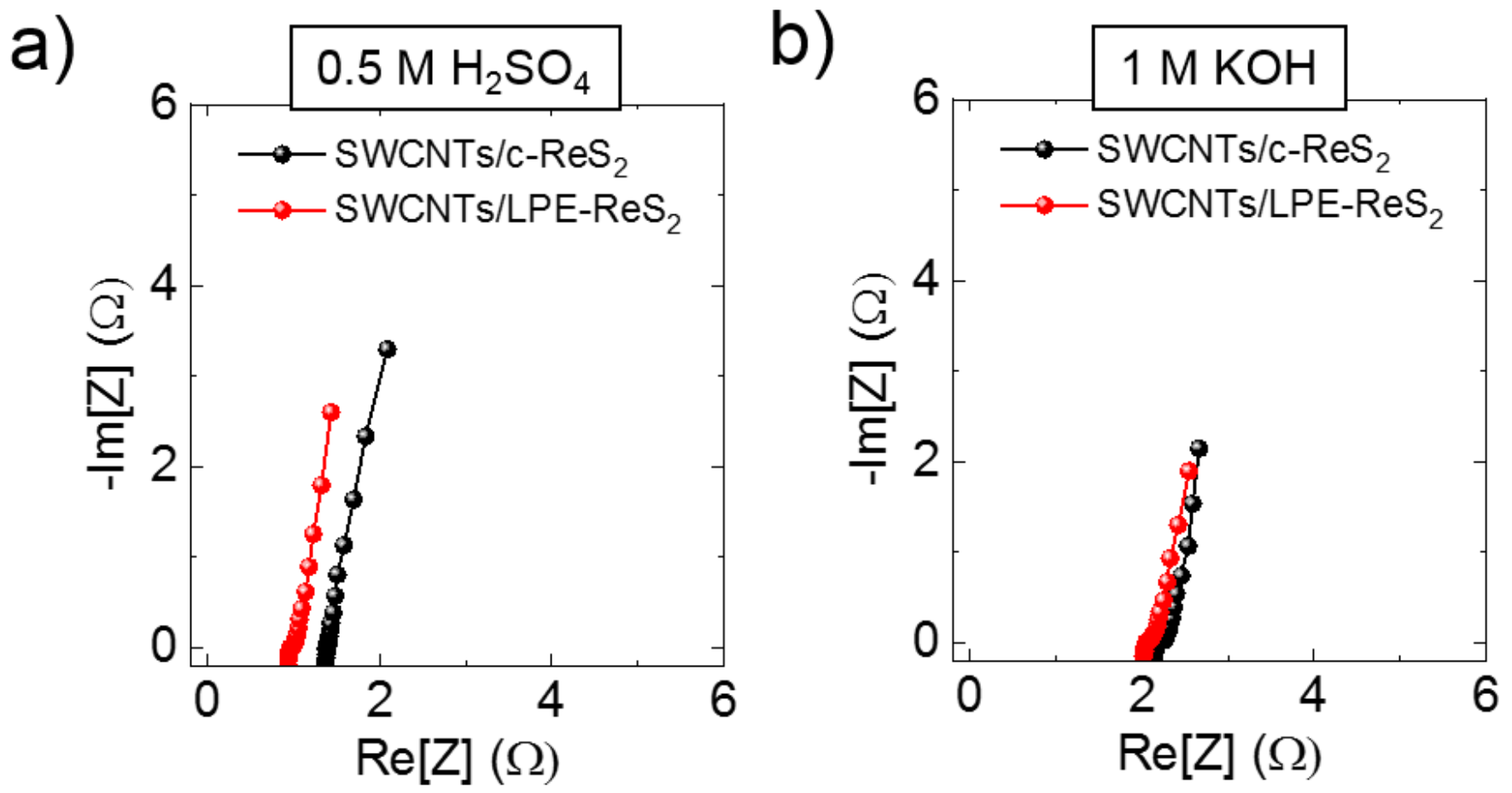

**Figure S11.** Nyquist plots SWCNTs/c-$ReS_2$ and SWCNTs/LPE-$ReS_2$ obtained by EIS measurements at open - circuit potential from 0.1 Hz to 200 kHz in (a) 0.5 M $H_2SO_4$ and (b) 1 M KOH.

Double-layer capacitance ($C_{dl}$) of the c-$ReS_2$ and LPE-$ReS_2$ films were estimated from cyclic voltammetry (CV) measurements in a non-Faradaic region of potential (potential between 0.30 and 0.45 V *vs.* Ag/AgCl) at various potential scan rates (ranging from 20 to 800 mV $s^{-1}$) in 0.5 M $H_2SO_4$. Double-layer capacitance provides information regarding the electrochemically accessible surface area of the electrode. Such physical quantity depends on both the physical surface area and the porosity of the electrode. The double-layer capacitances ($C_{dl}$) of the catalyst films were estimated. To estimate experimentally $C_{dl}$, the catalyst films were produced by drop casting the corresponding dispersions on flat substrates of glassy carbon (GC) (catalyst mass loading of 0.2 mg $cm^{-2}$). The use of flat GC as the substrate allows the $C_{dl}$ contribution of the substrate to be limited compared to the case of catalyst films deposited on SWCNTs (as for the electrodes investigated in the main text). By plotting the difference between the anodic and the cathodic current densities ($\Delta j = (j_a - j_c)$) at 0.375 V vs. RHE as a function of the scan rate (SR) (**Figure S12a**), the $C_{dl}$ was calculated from the slope of linear fit of Δj vs. SR, assuming $C_{dl} = (\Delta j)/2(SR)$. The calculated $C_{dl}$ of the c-$ReS_2$ film is ~0.162 mF $cm^{-2}$, which is ~1.5 time the one of the LPE-$ReS_2$ film. These results indicate that the electrochemical accessible surface area of c-$ReS_2$ film is superior to the LPE-$ReS_2$ film. This could be due to the small lateral size (~ 4±1 nm determined by TEM) and partial vertical orientation of the c-$ReS_2$ nanosheets relatively to the substrate creating a porous network as can be seen in the high magnification SEM image in **Figure S12b**, while the large LPE-$ReS_2$ flakes (> 100 nm determined by TEM) are randomly stacked, **Figure S12c**.

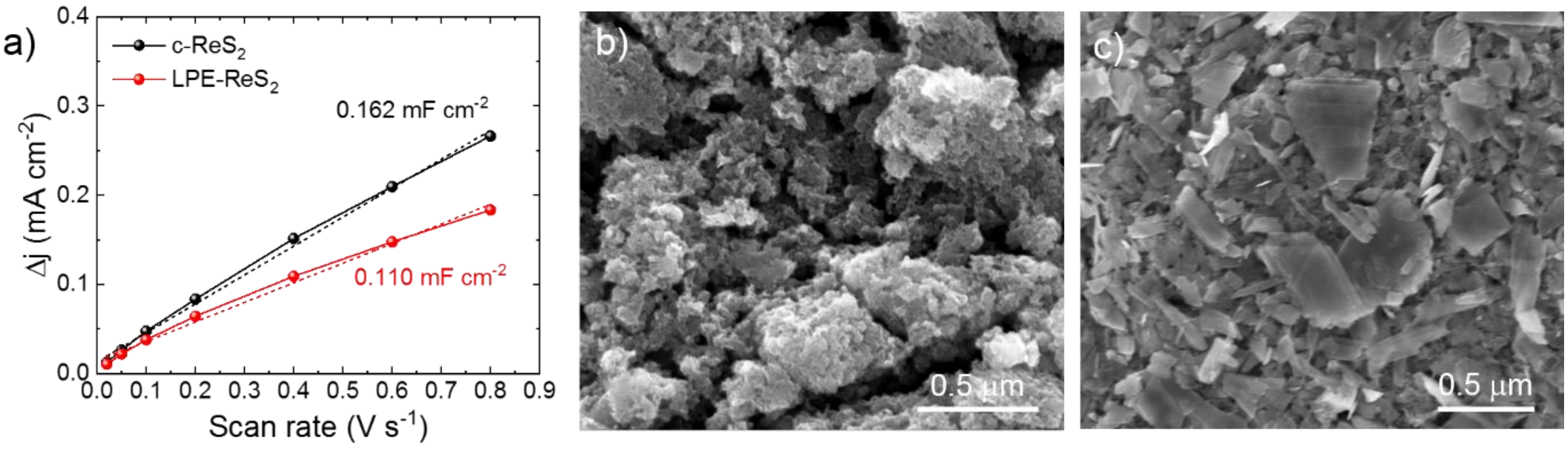

**Figure S12.** a) Scan rate dependence of the Δj calculated form CV curves of c-$ReS_2$ and LPE-$ReS_2$ films deposited on GC. The linear fits of the curves and the calculated $C_{dl}$ values are also shown. (b, c) Representative SEM images of the as-prepared SWCNTs/c-$ReS_2$ (b) and SWCNTs/LPE-$ReS_2$ (c) electrodes showing the arrangement of the c-$ReS_2$ nanosheets and LPE-$ReS_2$ flakes, respectively.

Moreover, we performed stability tests during 24 and 48 h in both acidic and alkaline solutions analyzing also how these conditions affect to the morphology of the SWCNTs/c-$ReS_2$ electrodes. These long electrochemical experiments induced the formation of cracks, but the SWCNTs prevented the electrodes from breaking (see **Figure S13** for acid and alkaline tests after 24h (a-f) and 48h (g-l)). As discussed in the main text, these changes are induced by the mechanical stresses caused by $H_2$ bubbling through the layered structure of the electrodes[16],[17] and active material reorganization during HER-activation.

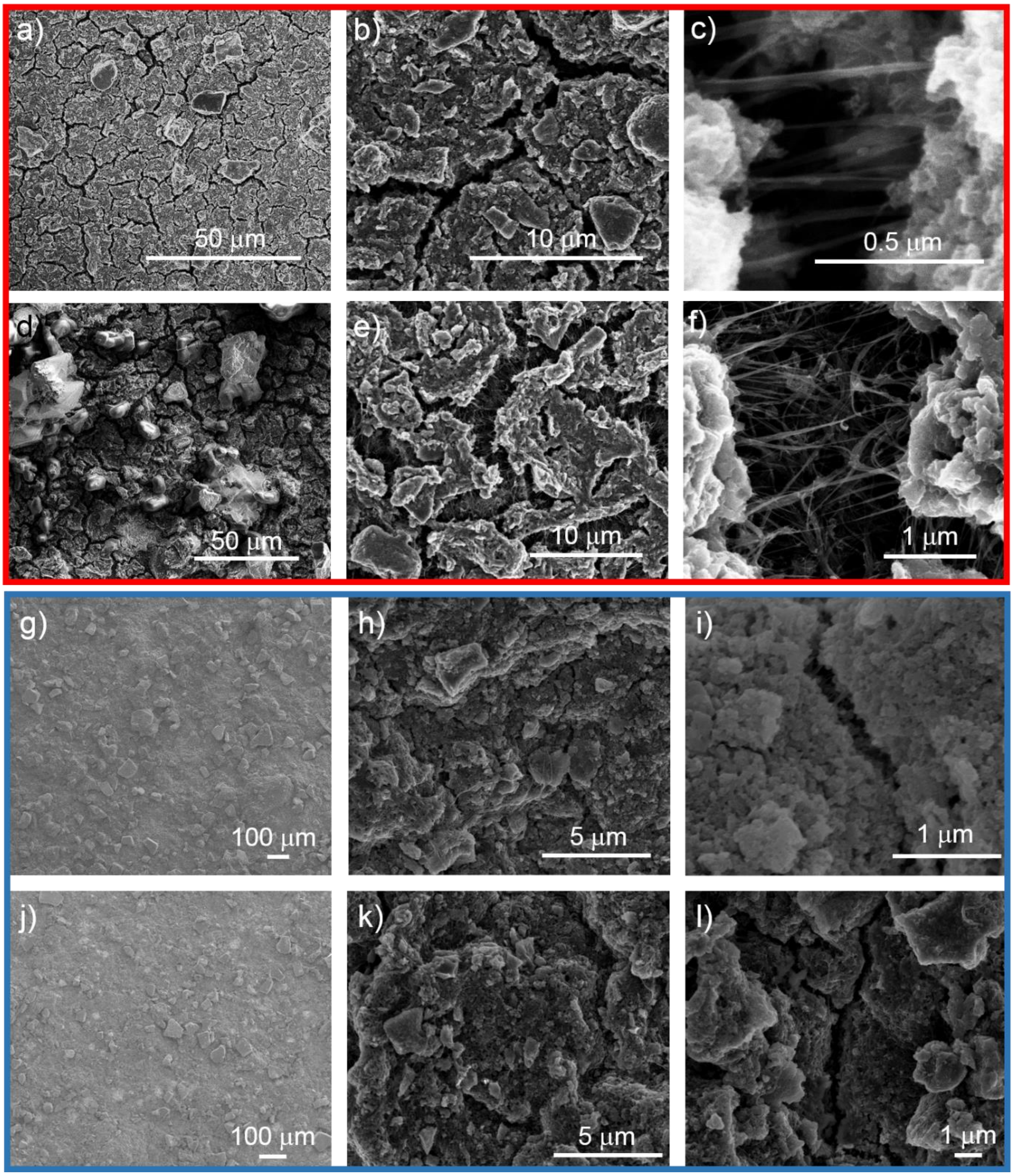

**Figure S13.** Top-view representative SEM images at different magnification of the SWCNTs/c-$ReS_2$ electrodes after stability test 24 h (red square, a-f) and 48 h (blue square, g-l) for HER in 0.5 M $H_2SO_4$ (a, b, c, g, h, i) (the enlargement increases from right panels to left panels) and in 1 M KOH (d, e, f, j, k, l) (the enlargement increases from right panels to left panels).

**Figure S14a-b** show the chronoamperometry measurements (current retention *vs.* time) for representative SWCNTs/c-$ReS_2$ (beyond those reported in the main text), over a 24 h in acidic and alkaline solutions, respectively. As mentioned in the main text, a constant overpotential was applied in order to provide the same starting cathodic current density of 20 mA $cm^{-2}$ for HER, and GC rod was used as counter-electrode. As shown in **Figure S14a**, in acidic solution, the electrode has shown a significant increase (of 10.7%) of the cathodic current density after 24 h. This increase of the activity in acidic condition can be ascribed to favourable morphological changes of the electrode film during the HER process, already observed in metallic TMDCs[18][19,20] or even in graphene-based electrocatalysts.[21] In alkaline solution (**Figure S14b**), the electrode retains ~79.5% of its initial cathodic current density. Despite the decrease of the HER-activity of the electrode observed in alkaline solutions, the iR-corrected polarization curves show that the c-$ReS_2$ nanosheets are electrochemically activated in both acidic and alkaline solutions after 24 h. In fact, $\eta_{10}$ is reduced from 0.195 V to 0.162 V in 0.5 M $H_2SO_4$, and from 0.327 V to 0.181 V in 1 M KOH. The decrease of the cathodic current density in alkaline solution is ascribed to the increase of the electrical resistance of the electrodes (from 4.4 Ω to 45.9 Ω). In fact, cracks are observed in the electrodes after the electrochemical measurements by SEM (**Figure S13a-f**). The morphological changes, which are more pronounced in alkaline solutions compared to those in acidic ones, can originate from both the mechanical stresses caused by $H_2$ bubbling through the layered structure of the electrodes[16],[17]. These stresses can cause a reorientation and/or fragmentation of the 2D electrocatalysts, as often observed in literature.[18–22] These morphological changes can be also responsible for the different electrochemical stability behavior of the HER activity observed for 24 h and 48 h tests (Figure 6c-d, main text). This can be avoided by using electrocatalyst binders, such as sulfonated tetrafluoroethylene-based fluoropolymer copolymers (*e.g.*, Nafion).

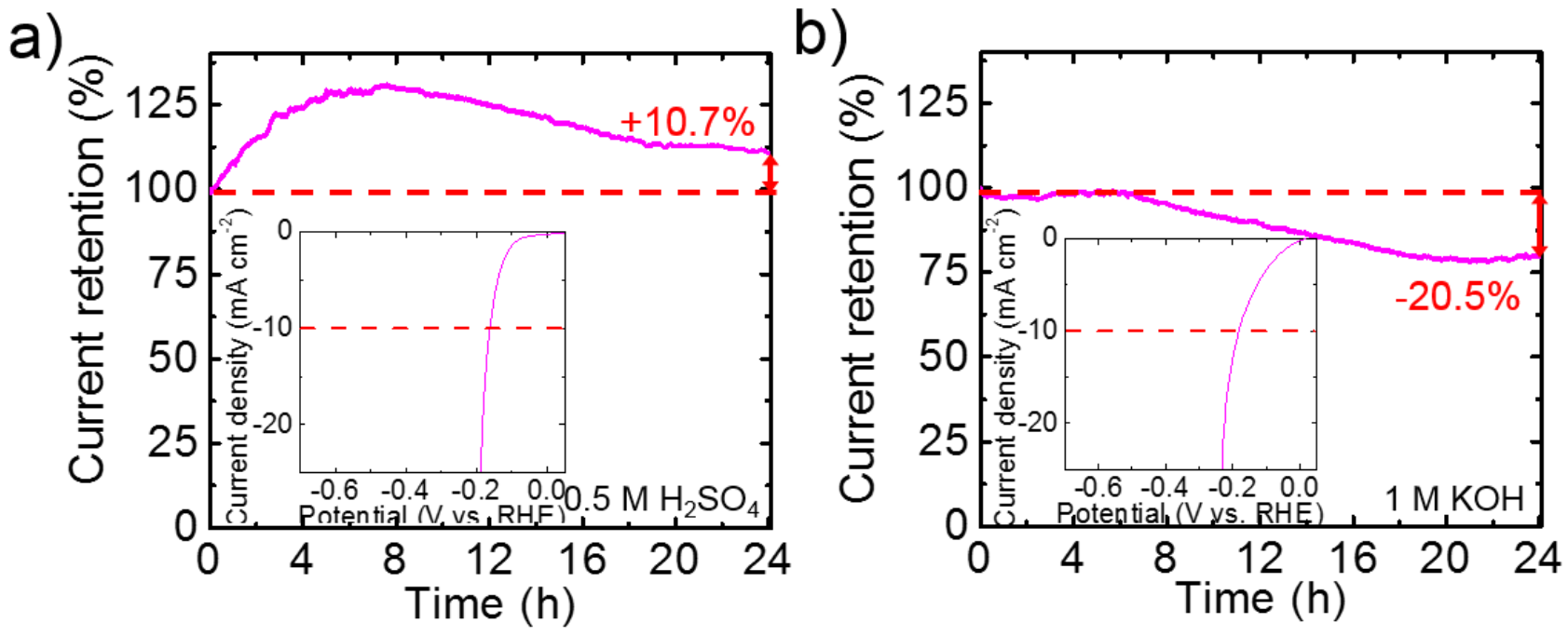

**Figure S14.** (a,b) Chronoamperometry measurements (current retention *vs.* time) for the SWCNTs/c-$ReS_2$ electrode in acidic (0.5 M $H_2SO_4$) and alkaline (1 M KOH) solutions. A constant overpotential has been applied in order to provid e an equal starting cathodic current density of 20 mA $cm^{-2}$. The inset panels of a) and b) show the iR-corrected polarization curves after stability tests.

## References:

[1] V. Nicolosi, M. Chhowalla, M. G. Kanatzidis, M. S. Strano, J. N. Coleman, *Science* **2013**, *340*, 1226419.

[2] F. Bonaccorso, A. Lombardo, T. Hasan, Z. Sun, L. Colombo, A. C. Ferrari, *Mater. Today* **2012**, *15*, 564.

[3] F. Bonaccorso, A. Bartolotta, J. N. Coleman, C. Backes, *Adv. Mater.* **2016**, 6136.

[4] J. Tauc, *Materials Research Bulletin* **1968**, *3*, 37.

[5] N. R. Pradhan, A. McCreary, D. Rhodes, Z. Lu, S. Feng, E. Manousakis, D. Smirnov, R. Namburu, M. Dubey, A. R. Hight Walker, H. Terrones, M. Terrones, V. Dobrosavljevic, L. Balicas, *Nano Letters* **2015**, *15*, 8377.

[6] A. McCreary, J. R. Simpson, Y. Wang, D. Rhodes, K. Fujisawa, L. Balicas, M. Dubey, V. H. Crespi, M. Terrones, A. R. Hight Walker, *Nano Letters* **2017**, *17*, 5897.

[7] M. Hafeez, L. Gan, H. Li, Y. Ma, T. Zhai, *Advanced Materials* **2016**, *28*, 8296.

[8] S. Jiang, M. Hong, W. Wei, L. Zhao, N. Zhang, Z. Zhang, P. Yang, N. Gao, X. Zhou, C. Xie, J. Shi, Y. Huan, L. Tong, J. Zhao, Q. Zhang, Q. Fu, Y. Zhang, *Communications Chemistry* **2018**, *1*, 17.

[9] D. Chen, Y. Gao, Y. Chen, Y. Ren, X. Peng, *Nano Letters* **2015**, *15*, 4477.

[10] Z. Li, N. Dong, Y. Zhang, J. Wang, H. Yu, F. Chen, *APL Photonics* **2018**, *3*, 080802.

[11] D. Wolverson, S. Crampin, A. S. Kazemi, A. Ilie, S. J. Bending, *ACS Nano* **2014**, *8*, 11154.

[12] G. Gouadec, P. Colomban, *Progress in Crystal Growth and Characterization of Materials* **2007**, *53*, 1.
[13] S. Yang, C. Jiang, S. Wei, *Applied Physics Reviews* **2017**, *4*, 021304.
[14] "Advanced Specialty Gas Equipment (ASGE)," can be found under http://www.asge-online.com/pdf/ASGEpg185.pdf, **n.d.**
[15] S. Yang, J. Kang, Q. Yue, J. M. D. Coey, C. Jiang, *Advanced Materials Interfaces* **2016**, *3*, 1500707.
[16] L. Najafi, S. Bellani, R. Oropesa-Nuñez, A. Ansaldo, M. Prato, A. E. Del Rio Castillo, F. Bonaccorso, *Advanced Energy Materials* **2018**, *8*, 1703212.
[17] L. Najafi, S. Bellani, R. Oropesa-Nuñez, A. Ansaldo, M. Prato, A. E. Del Rio Castillo, F. Bonaccorso, *Advanced Energy Materials* **2018**, *8*, 1801764.
[18] Y. Liu, J. Wu, K. P. Hackenberg, J. Zhang, Y. M. Wang, Y. Yang, K. Keyshar, J. Gu, T. Ogitsu, R. Vajtai, J. Lou, P. M. Ajayan, B. C. Wood, B. I. Yakobson, *Nature Energy* **2017**, *2*, 17127.
[19] J. Zhang, J. Wu, X. Zou, K. Hackenberg, W. Zhou, W. Chen, J. Yuan, K. Keyshar, G. Gupta, A. Mohite, P. M. Ajayan, J. Lou, *Materials Today* **2019**, *25*, 28.
[20] L. Najafi, S. Bellani, R. Oropesa-Nuñez, B. Martin-Garcia, M. Prato, V. Mazanek, D. Debellis, S. Laucello, R. Brescia, Z. Sofer, F. Bonaccorso, *J. Mater. Chem. A* **2019**, *just accepted*, DOI 10.1039/c9ta07210a.
[21] L. Najafi, S. Bellani, R. Oropesa-Nunez, B. Martin-Garcia, M. Prato, F. Bonaccorso, *ACS Applied Energy Materials* **2019**, *2*, 5373.
[22] J. Shi, X. Wang, S. Zhang, L. Xiao, Y. Huan, Y. Gong, Z. Zhang, Y. Li, X. Zhou, M. Hong, Q. Fang, Q. Zhang, X. Liu, L. Gu, Z. Liu, Y. Zhang, *Nature Communications* **2017**, *8*, DOI 10.1038/s41467-017-01089-z.